\documentclass[12pt]{iopart}
\usepackage{graphicx,color}
\usepackage{comment}
\usepackage{textcomp}

\begin{document}

\newcommand {\nc} {\newcommand}
\nc {\IR} [1]{\textcolor{red}{#1}}
\nc {\IB} [1]{\textcolor{blue}{#1}}

\title{Quantifying Uncertainties on Fission Fragment Mass Yields With Mixture Density Networks}

\author{A.E. Lovell, A.T. Mohan, and P. Talou}
\address{Los Alamos National Laboratory, Los Alamos, NM 87545, USA}
\ead{lovell@lanl.gov, arvindm@lanl.gov}

\vspace{10pt}
\begin{indented}
\item[] \today
\end{indented}

\begin{abstract}
Probabilistic machine learning techniques can learn both complex relations between input features and output quantities of interest as well as take into account stochasticity or uncertainty within a data set.  In this initial work, we explore the use of one such probabilistic network, the Mixture Density Network (MDN), to reproduce fission yields and their uncertainties.  We study mass yields for the spontaneous fission of $^{252}$Cf, exploring the number of training samples needed for converged predictions, how different levels of uncertainty propagate from the training set to the MDN predictions, and how well physical constraints of the yields - such as normalization and symmetry - are upheld by the algorithm.  Finally, we test the ability of the MDN to interpolate between and extrapolate beyond samples in the training set using energy-dependent mass yields for the neutron-induced fission on $^{235}$U.  The MDN provides a reliable way to include and predict uncertainties and is a promising path forward for supplementing sparse sets of nuclear data.
\end{abstract}

\vspace{2pc}


\section{Motivation}
\label{sec:mot}

Completely and accurately describing the nuclear fission process from the formation of the compound fissioning nucleus (or the initial conditions for spontaneously fissioning nuclei) to the emission of prompt and delayed particles requires a range of separate models.  Microscopic and macroscopic-microscopic models can describe how a compound nucleus deforms and scissions and statistical models describe the evaporation of neutrons and $\gamma$ rays from the initial fission fragments.  Among the many quantities needed as input and to validate models of the later stage of fission are the yields of the fission fragments, in mass, charge, kinetic or excitation energy, spin, and parity before prompt particle emission.  For several major actinides, many of these yields have been well studied experimentally, in particular for spontaneous fission of $^{252}$Cf. For neutron-induced fission of major and minor actinides, most of the studies have been performed using thermal neutrons, and as the neutron energy increases, measurements become more scarce.  The experimental measurements become more difficult as the incident neutron energy increases and when multiple inputs are required to be measured simultaneously, often due to the limited nature of the experimental setup or complicated backgrounds.  In addition, several of these inputs are difficult or impossible to measure directly, such as the spin distribution of the initial fission fragments.

Microscopic models based on nucleon-nucleon interactions can make predictions for a number of these quantities, in particular joint mass and charge yields - and in some cases the kinetic energy of each fragment, e.g. \cite{Schunck2014,Regnier2016}.  Although these models can be applied to fissile nuclei across the nuclear chart, the predictions are not necessarily accurate. The same is true of microscopic-macroscopic models \cite{Sierk2017,Verriere2019,Moller2015} which typically model the nucleus as a liquid drop and add microscopic corrections to that prescription.    Where fundamental calculations are not of high-enough quality to make predictions for these inputs, systematics have been developed.

These types of systematics have been developed for many inputs, particularly those that are directly measurable, such as the mass, charge, and kinetic energy distributions of the fission fragments.  A.C. Wahl constructed systematics for both mass and charge yields for the actinide region \cite{Wahl2002}.  Often, mass yields are parameterized as a sum of Gaussian distributions, as in \cite{CGMF,Becker2013,FREYA1,FREYA2,BeoH}, and the same can be done for the total kinetic energy yields for each fragment mass.  The weights, means, and standard deviations of each of these Gaussian distributions are typically taken to be function of the incident neutron energy so that parameterizations can be interpolated between and extrapolated beyond the data used to constrain the parametrizations.  A slightly more complex shape parameterization for joint mass-kinetic energy yield was developed by Brosa \cite{Brosa1990}.  Spin distributions are often assumed to be Gaussian \cite{CGMF,BeoH}, with a width that can depend on the incident energy.  In all cases, these parameterizations need to be fit to experimental data or theoretical simulations. 

To benchmark any of these methods or constrain the fitting parameters, experimental data are needed for multiple targets and at various incident neutron energies.  In addition, to validate models that compute joint distributions, it is useful to measure many fission fragment properties simultaneously.  With many current experimental setups, this is not the case \cite{Devlin2018,Pozzi2014,Ullmann2013,Pellereau2017,Meierbachtol2015,Fregeau2016,Heffner2014}, so constraining any joint distribution requires using data from different measurements or data only containing a part of the distribution.  For example, a mass yield may be measured for a nucleus at a given energy but for that same reaction, only the average kinetic energy of each fragment is known and not the full distribution of kinetic energy for each mass. Novel methods are required to fill in this sparse data landscape. In addition, the errors and uncertainties that are inherently associated with the experimental measurements should be taken into account and propagated to the observable of interest.

In nuclear physics, several groups are beginning to explore machine learning techniques to interpolate and extrapolate models with well-quantified uncertainties.  Gaussian processes and feed-forward neural networks have been used to build emulators for models of heavy ion collisions \cite{Novak2014} and nuclear properties \cite{Regnier2019} that can be used to more quickly perform parameter optimizations.  In addition, there have been some prominent uses of Gaussian processes for uncertainty quantification, particularly in determining the neutron dripline of nuclei far from stability \cite{Neufcourt2018,Neufcourt2019}.  Bayesian Neural Networks, in which a Bayesian analysis is performed on a deterministic-map neural network, have recently become popular as well, in predicting residuals of mass models \cite{Utama2016,Neufcourt2018} and fission yield models \cite{Wang2019}.  However, both Gaussian processes and Bayesian Neural Networks have their shortcomings.  Because the Gaussian process emulators return to their mean values outside of the training region, these cannot be used to reliably extrapolate beyond the training set, and the Bayesian Neural Networks are extremely computationally demanding, having to retrain the network possibly thousands of times to perform the full analysis.

Here, we propose to use a machine learning algorithm, the Mixture Density Network (MDN) \cite{MDN}, which can take into account uncertainties and correlations within the training data set and propagate these uncertainties to predicted quantities.  While many other machine learning techniques provide only a mean value and variance of the mean, the MDN provides a unique advantage since it estimates the full posterior distribution, instead of having to assume a shape for this distribution.  This is particularly powerful when dealing with discrepant data sets in the training procedure, as they are not simply averaged together.

In this paper, we provide an initial study of the MDN as applied to fission fragment mass yields.  We first describe the Mixture Density Network in Section \ref{sec:MDN}.  In Section \ref{sec:convergence}, we show studies for the number of training points that are required for converged results as well as a study of how the uncertainties propagate from the training set to the testing set using $^{252}$Cf(sf).  Finally, in Section \ref{sec:interp}, we show how the MDN can be used to interpolate between data sets using both simulated models and experimental data with the energy dependence of $^{235}$U(n,f) and conclude in Section \ref{sec:outlook}.

\section{Mixture Density Network}
\label{sec:MDN}

In a standard feed-forward neural network (NN), the aim is to optimize a complex, non-linear mapping, $y=f(x)$, between an input, $x$, and an output, $y$.  The elementary unit of the NN is an artificial neuron (or node), which has a weight and a bias.  The nodes are arranged in layers and these layers together form the NN.  The learning process optimizes the weights and biases for each node to map the input to the output typically through maximum likelihood estimation (MLE) based loss functions.  Part of developing the network is determining the optimal hyper-parameters, i.e. the number of layers and nodes per layer, to perform this mapping.  Often, this is done through trial and error; while a deeper network will generally give a better agreement between the training set and the output of the neural network, one must be careful to avoid overfitting:  where the network exactly maps the input to the output but does not generalize to the mapping $f$, therefore having no predictive power outside of the training dataset. While such MLE-based algorithms are successful in a variety of applications where the input-output mapping is deterministic, they struggle to learn input-output mappings that are strongly probabilistic.  Such situations are often encountered in nuclear physics - and data science more widely - where each measurement has an associated uncertainty (as well as discrepancies between measurements), and the machine learning methods that are typically discussed in nuclear physics such as Gaussian processes and Bayesian neural networks have shortcomings, as discussed in the introduction. 

To mitigate these challenges, we propose using a novel machine learning technique, the Mixture Density Network (MDN) \cite{MDN}.  Although this method was proposed in the early 1990's, few applications of it exist in scientific research, e.g. \cite{MDNRibes,MDNHerzallah,MDNZen,Disanto}.  Unlike standard neural networks which deterministically map the input to the output using MLE, the MDN instead describes the output probabilistically as a sum of Gaussian distributions,
\begin{equation}
y(\textbf{x}) = \sum \limits _{i=1} ^m \alpha_i(\textbf{x}) \mathcal{N}\left [\mu_i(\textbf{x}),\sigma_i(\textbf{x}) \right ],
\label{eqn:MDNGauss}
\end{equation}

\noindent where the weights, means, and standard deviations, $\alpha_i$, $\mu_i$, and $\sigma_i$, are determined by a feed-forward neural network and $\textbf{x}$ is the vector of input features.  The loss function is also no longer a simple MLE, but  a log-likelihood loss,
\begin{equation}
\mathcal{L} = - \ln {\left [ \sum \limits _{i=1} ^m \frac{\alpha_i(\textbf{x})}{(2\pi)^{m/2} \sigma_i (\textbf{x})} \mathrm{exp} \left \{ -\frac{|| \textbf{t}-\mu_i(\textbf{x}) ||^2}{2\sigma_i (\textbf{x})^2} \right \} \right ]},
\label{eqn:logloss}
\end{equation}

\noindent where $\textbf{t}$ is the vector of training outputs and $m$ is the total number of Gaussian mixtures.  Equation (\ref{eqn:logloss}) minimizes the difference of probability distributions of the true and predicted posteriors rather than the absolute root mean square error, which is more suitable for deterministic datasets.  In essence, the MDN parameterizes the mixture of Gaussians based on data, rather than using empirical estimates.

The MDN provides several benefits over standard, deterministic-map neural networks.  Because the weights, means, and standard deviations are predicted by the NN instead of the output directly, draws made from Eq. (\ref{eqn:MDNGauss}) form a posterior distribution for each predicted value.  Therefore, instead of providing the mean and standard deviation of each predicted value and having to assume the resulting distribution, the MDN attempts to directly learn the exact posterior distribution.  The gives a more consistent interpretation of the result of including discrepant data sets within a total training set; instead of averaging the two sets to make a prediction like MLE, the MDN can form a multi-modal posterior distribution over both jointly. More details on the differences between deterministic and probabilistic approaches, like the MDN, can be found in \cite{MDN}.

We implemented our MDN in \texttt{PyTorch} \cite{pytorch} to run on both CPUs and GPUs - although there is a significant speed up in the computation time when running on GPUs.  The formulation of the MDN is general enough to allow for a change in the network architecture:  for any run, the number of layers and nodes in each layer can be easily specified.  In addition, there is no limit to the number of input features that can be included or the number of Gaussian mixtures, though it may increase computational costs.  Although in this work, we only look at networks that map a one input to one output or many inputs to one output, the architecture is flexible enough that it can also handle mapping to many outputs; currently for this case, the number of Gaussian mixtures is the same for each output.


\section{Convergence studies}
\label{sec:convergence}

To begin, we first perform several numerical convergence studies.  Although, machine learning algorithms tend to be sensitive to the hyper-parameters of the neural network  - that is the numbers of layers and nodes - as well as the size and scope of the training set, we have fixed the network hyper-parameters and concentrate on the properties of the training set.  By doing so, we can attribute any significant changes in the MDN output to changes in the input.  In this work, we have optimized the size of the NN to best reproduce the training set in these initial studies with the smallest number of nodes and layers.  We focus our studies primarily on how the uncertainties propagate from the training set to the MDN predictions, as we first want to understand to what extent the MDN incorporates uncertainties in a meaningful way.

We begin our study with mass yields, $Y(A)$, for the spontaneous fission of $^{252}$Cf.  $^{252}$Cf(sf) provides an ideal test case because there is a large amount of well-measured experimental data for mass yields \cite{Hambsch1997,Gook2014,BudtzJorgensen1988,Romano2010,Zeynalov2011,Kozulin2011} but also for most other prompt observables, e.g. \cite{Talou2018} and references therein.  These mass yields can be well described by a sum of several Gaussians, e.g. \cite{CGMF,Becker2013,BeoH}, giving a straightforward way to mock up uncertainties on the distribution and then study how these uncertainties propagate from the training set to the resulting MDN prediction.

Unless otherwise noted, the $Y(A)$ training sets in this section are constructed from a three Gaussian parameterization,
\begin{equation}
\label{eqn:threeGaussian}
Y(A|E_\mathrm{inc}) = G_0 (A|E_\mathrm{inc}) + G_1 (A|E_\mathrm{inc}) + G_2 (A|E_\mathrm{inc}),
\end{equation}

\noindent with
\begin{equation}
G_0(A|E_\mathrm{inc}) = \frac{W_0(E_\mathrm{inc})}{\sqrt{2\pi} \sigma_0(E_\mathrm{inc})}\exp  \left [ \frac{-(A-A_c/2)^2}{2\sigma_0 (E_\mathrm{inc})^2} \right ],
\end{equation}

\noindent and
\begin{eqnarray}
G_{1,2} (A|E_\mathrm{inc}) & = \frac{W_{1,2}(E_\mathrm{inc})}{\sqrt{2\pi} \sigma_{1,2}(E_\mathrm{inc})} \left \{ \exp \left [ \frac{-(A-\mu_{1,2}(E_\mathrm{inc}))^2}{2\sigma _{1,2} (E_\mathrm{inc})^2} \right ]  \right. \\
& \left. + \exp \left [ \frac{-(A-(A_c-\mu_{1,2}(E_\mathrm{inc})))^2}{2\sigma _{1,2} (E_\mathrm{inc})^2} \right ] \right \},
\end{eqnarray}

\noindent where $A_c$ is the mass number of the fissioning compound nucleus.  The functional forms of $W_i(E_\mathrm{inc})$, $\mu_i(E_\mathrm{inc})$, and $\sigma_i(E_\mathrm{inc})$ have been determined and fitted elsewhere to experimental data \cite{CGMF}.  The training set is then sampled from the distribution
\begin{equation}
\mathcal{N}[Y(A|E_\mathrm{inc}),\delta Y(A|E_\mathrm{inc})],
\label{eqn:trainingSampling}
\end{equation}

\noindent where $\mathcal{N}$ is a normal distribution and for testing purposes, in this work, $\delta$ varies between 0.01 and 0.20 (1\% to 20\%).  We could replace $\delta$ with a full correlation matrix between the masses, as we know the mass distribution is highly correlated.  A full correlation matrix would introduce more physics constraints in the machine learning algorithm.  We take a first step towards including these types of constraints in Sec. \ref{sec:constraints}, but an inclusion of the full correlation between masses is beyond the scope of this work.  

The training sets are sampled from Eq. (\ref{eqn:trainingSampling}) in the range $[A_c/2-36,A_c/2+36]$.  This limited range of fragment masses excludes mass yields less than $\sim 10^{-3}$, a region that causes the MDN training to become unstable due to the several orders of magnitude spanned by the tails of the distributions (to facilitate the training of the network, the training set is scaled linearly such that the mean is 0 and the standard deviation is 1, which is standard practice in machine learning and optimization).  These instabilities could be mitigated by using a logarithmic scaling of the training set, but that extension is beyond the scope of this work.  To test the resulting MDN simulations, we pull 10,000 samples from the Gaussian mixture within the same mass range; this large number of samples ensures that the constructed distributions are statistically significant.

For all of the test cases in Sec. \ref{sec:trainingPoints} and \ref{sec:uncertainty}, we use a single Gaussian mixture in Eq. (\ref{eqn:MDNGauss}).  Although a single Gaussian would appear to counteract the benefits of using an MDN in the first place, because our training set is sampled from a single Gaussian at each mass value, a single Gaussian mixture should be all that is needed to reproduce the set.  Too many Gaussians in the MDN leads to an increase in computational cost, since each mixture adds to the number of parameters in the NN.  The MDN neural network then consists of a single layer with 10 nodes.  Changes to the number of Gaussian mixtures when the experimental data are taken into account are discussed in Sec. \ref{sec:experimentalData}.  The same network architecture was used for the energy interpolation and extrapolations of Sec. \ref{sec:interp}.

\subsection{Convergence on the number of training points}
\label{sec:trainingPoints}

\begin{figure}
\centering
\begin{tabular}{cc}
\includegraphics[width=0.5\textwidth]{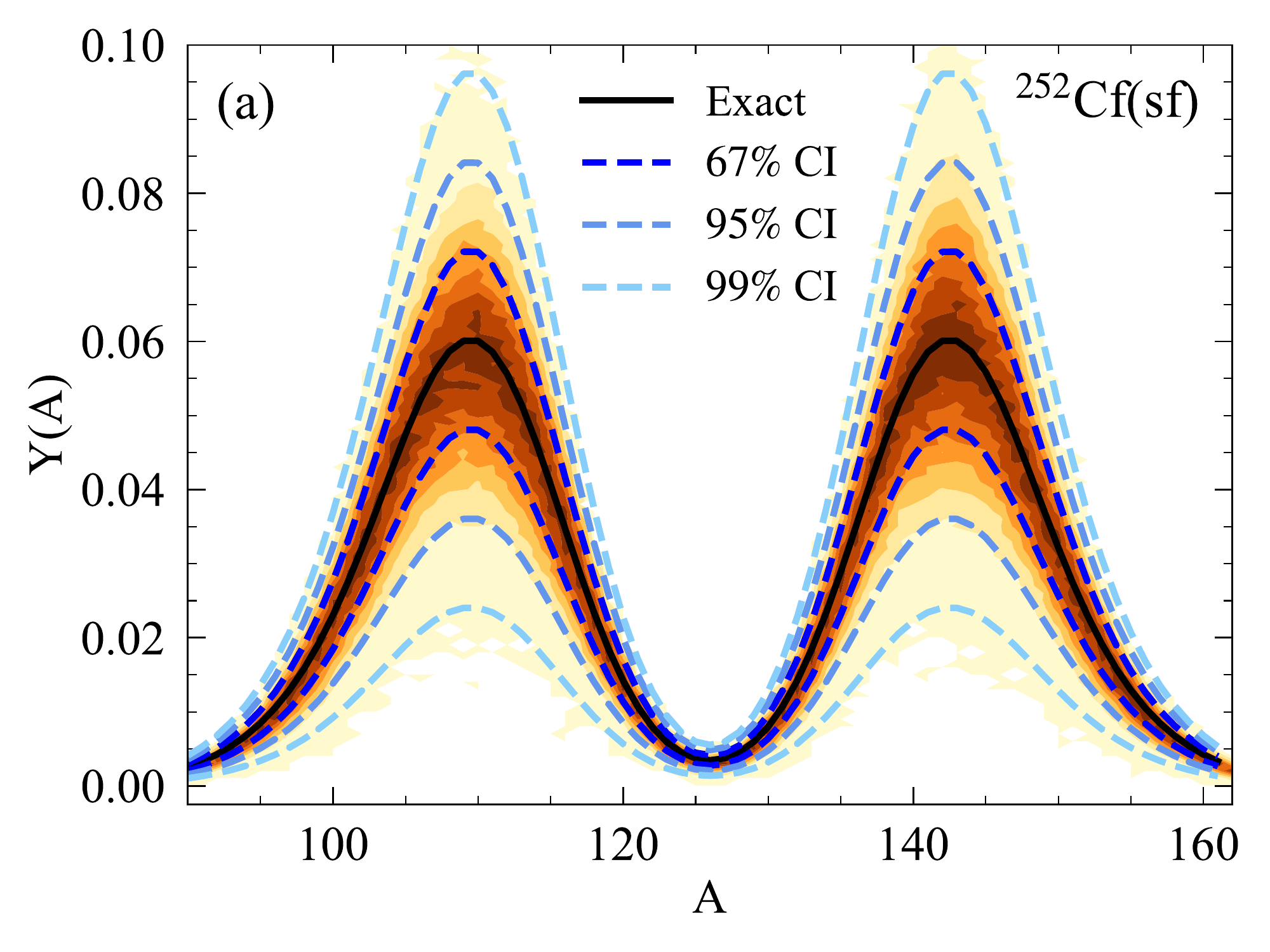} & \includegraphics[width=0.5\textwidth]{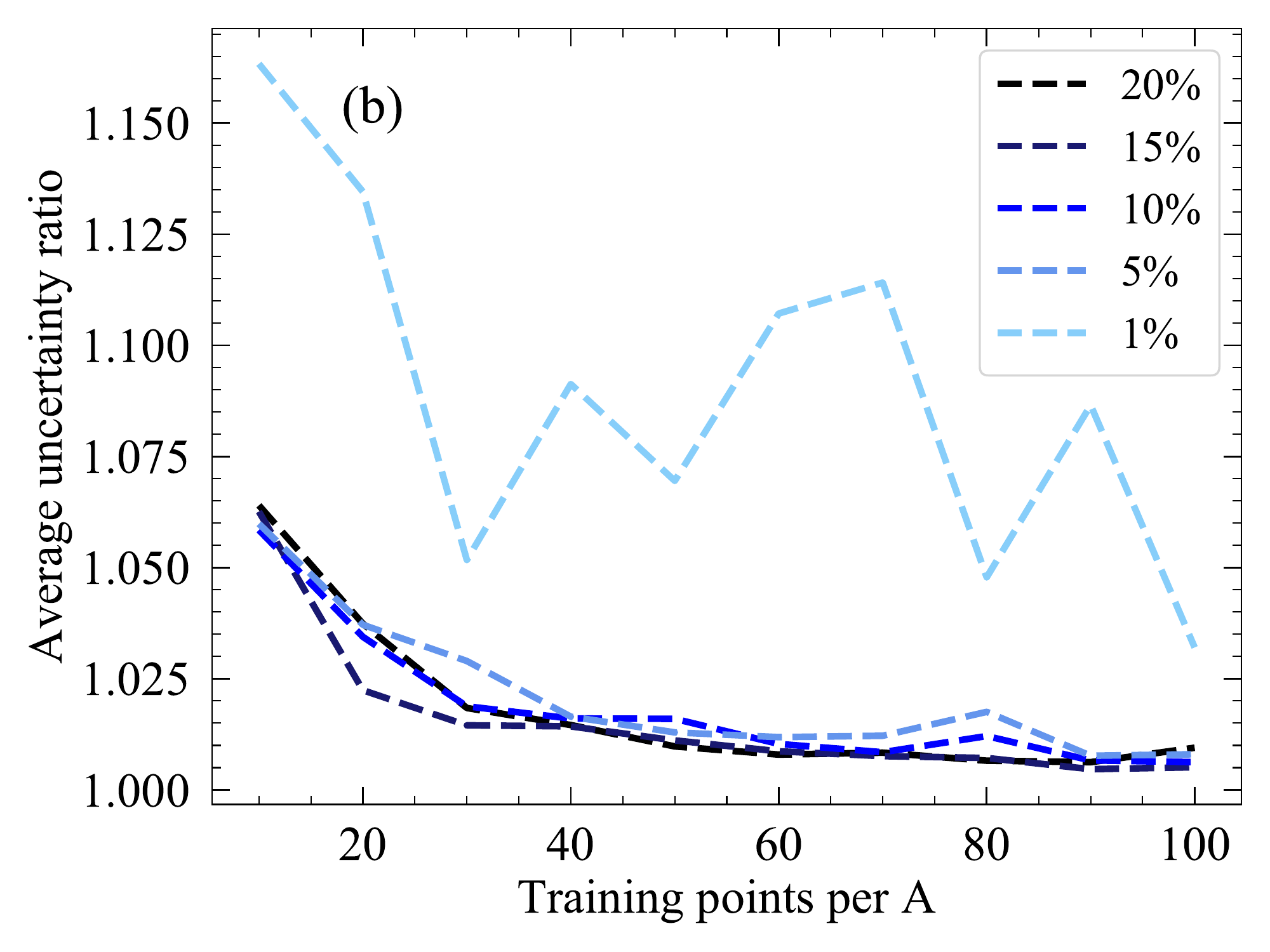}
\end{tabular}
\caption{(color online) (a) Density of the MDN prediction for the mass yields for $^{252}$Cf compared to the $Y(A)$ distribution from \texttt{CGMF}, black, and the $1\sigma$, $2\sigma$, and $3\sigma$ confidence intervals for 20\% uncertainty, blue dashed.  (b) The ratio of average uncertainty, $\varepsilon_\mathrm{MDN}/\varepsilon_\mathrm{training}$ as defined in Eq. (\ref{eqn:averageUncertainty}), for $20\%$, $15\%$, $10\%$, $5\%$, and $1\%$ uncertainties are included on the training set.}
\label{fig:ndataConvergence}
\end{figure}

To test the convergence of the number of samples needed in the training set for each mass value, we begin with $\delta=0.20$ using mass yields for $^{252}$Cf(sf).  While this uncertainty is much larger than the experimentally measured uncertainties, if the results with the largest uncertainty are converged, training sets with smaller uncertainties should also converge with the same sample size.  In Fig. \ref{fig:ndataConvergence}(a), we show the density distribution of the MDN prediction for $Y(A)$ of $^{252}$Cf(sf) when 50 samples per mass are included in the training set compared to the mean value of $Y(A)$ from the training set, black dashed line.  The densest part of the MDN distribution (dark orange) overlaps with the mean $Y(A)$ distribution (black solid line).  In addition, the blue dashed lines outline the $1\sigma$, $2\sigma$, and $3\sigma$ confidence interval for 20\% uncertainties on the mean value, which capture the highest density of the MDN predictions.

Because we are interested in how the uncertainties propagate from the training set to the MDN predictions, we quantify the agreement between uncertainties on the training set and the MDN results by plotting the average uncertainty, defined as
\begin{equation}
\varepsilon = \frac{1}{N} \sum \limits _{i=1} ^{N} \frac{\sigma(A_i)}{\overline{Y}(A_i)},
\label{eqn:averageUncertainty}
\end{equation}

\noindent where $\sigma(A_i)$ is the standard deviation of the MDN mass yield for mass $A_i$, $\overline{Y}(A_i)$ is the average of the MDN mass yields for mass $A_i$, and $N$ is the number of masses in the distribution.  In Fig. \ref{fig:ndataConvergence}(b), we plot $\varepsilon_\mathrm{MDN}/\varepsilon_\mathrm{training}$ for $\delta=0.20$ in black.  When at least 50 points per mass are included in the training set, the ratio between $\varepsilon_\mathrm{MDN}$ and $\varepsilon_\mathrm{training}$ has flattened out - indicating convergence of the number of samples included in the training set, even though this ratio does not reached one (a perfect agreement between the training set and the prediction).  


Also included in Fig. \ref{fig:ndataConvergence}(b), are the ratios of $\varepsilon_\mathrm{MDN}/\varepsilon_\mathrm{training}$ for $\delta=0.15$, $0.10$, $0.05$, and $0.01$; 1\% uncertainties are smaller than would be expected on the experimental data, but in this way, we can probe the extreme values that we might face.  When $15\%$, $10\%$, and $5\%$ uncertainties are considered, the convergence pattern is similar to that for $20\%$ uncertainties.  With $1\%$ uncertainties on the training set, the ratio of average uncertainties does trend towards 1 as more samples are included in the training set however, there are many more fluctuations, and the ratio is never as close to 1 as for the other uncertainty levels.  Therefore, when the uncertainties on the training set are small, we should be careful in interpreting our results from the MDN.

In practice, we do not expect that we will always have at least 50 experimental data points for each input to include in the training set, especially if the focus is on a single observable for a single isotope or in a restricted region of the nuclear chart.  However, we can always include an arbitrary number of data points by assuming that the experimental errors are Gaussian and sampling training points from $\mathcal{N}(d_i,\sigma_{d_i})$ where $d_i$ are the data points and $\sigma_{d_i}$ are the experimental errors, in the same way as we have sampled the training set for the convergence studies.

\subsection{Convergence of the uncertainty on the training set}
\label{sec:uncertainty}

\begin{figure}
\centering
\begin{tabular}{ccc}
\includegraphics[width=0.33\textwidth]{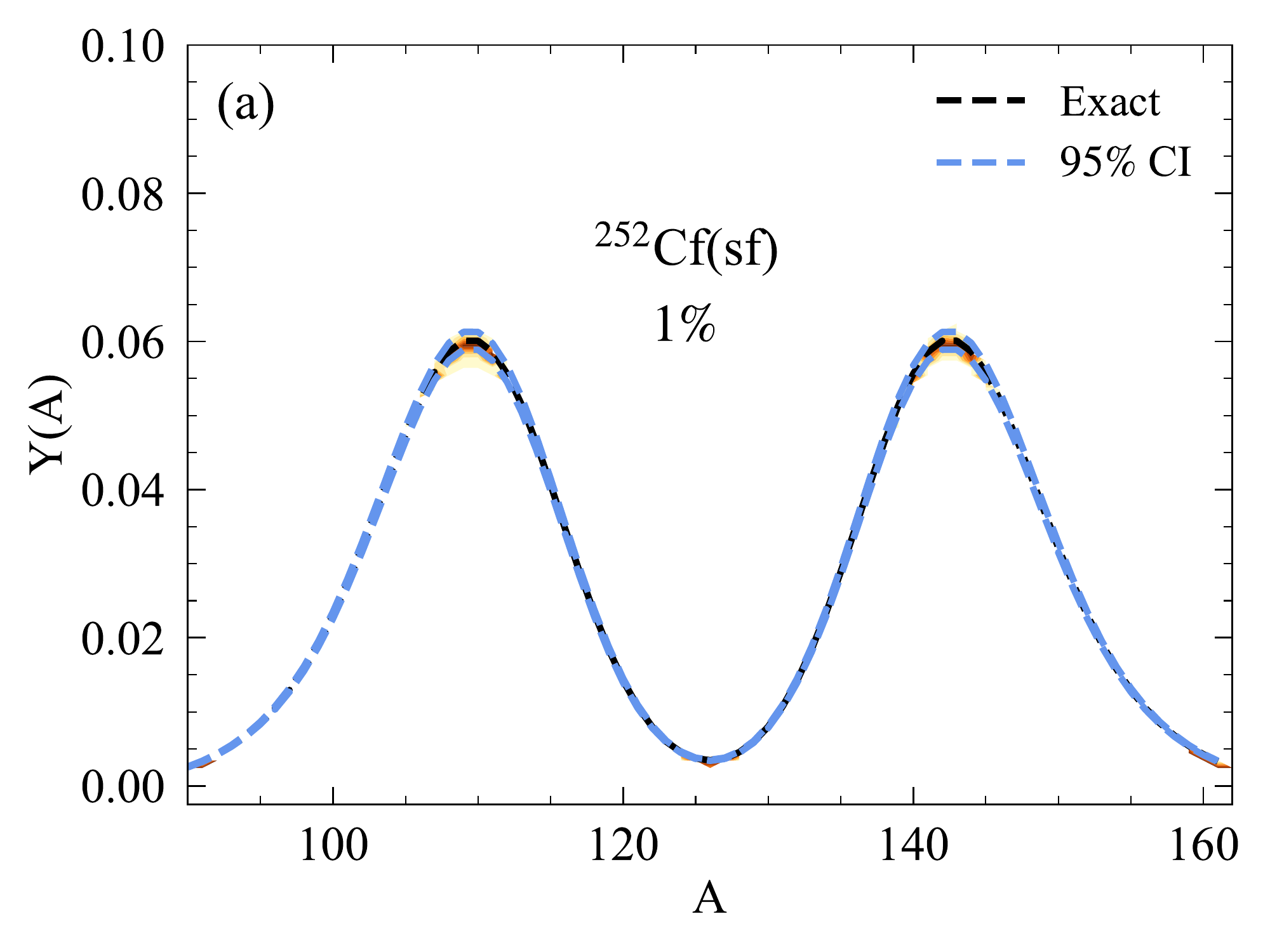} & \includegraphics[width=0.33\textwidth]{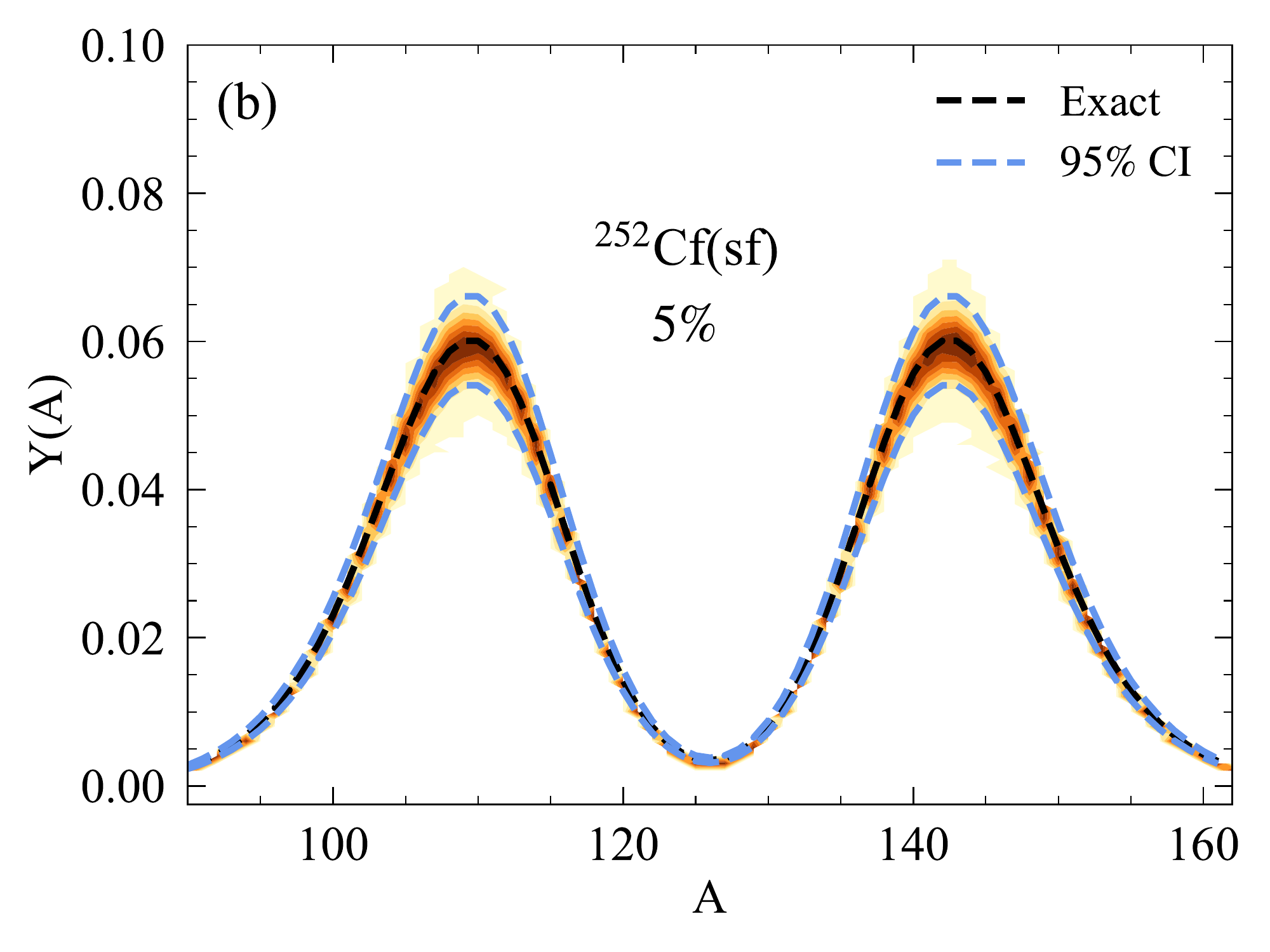} & \includegraphics[width=0.33\textwidth]{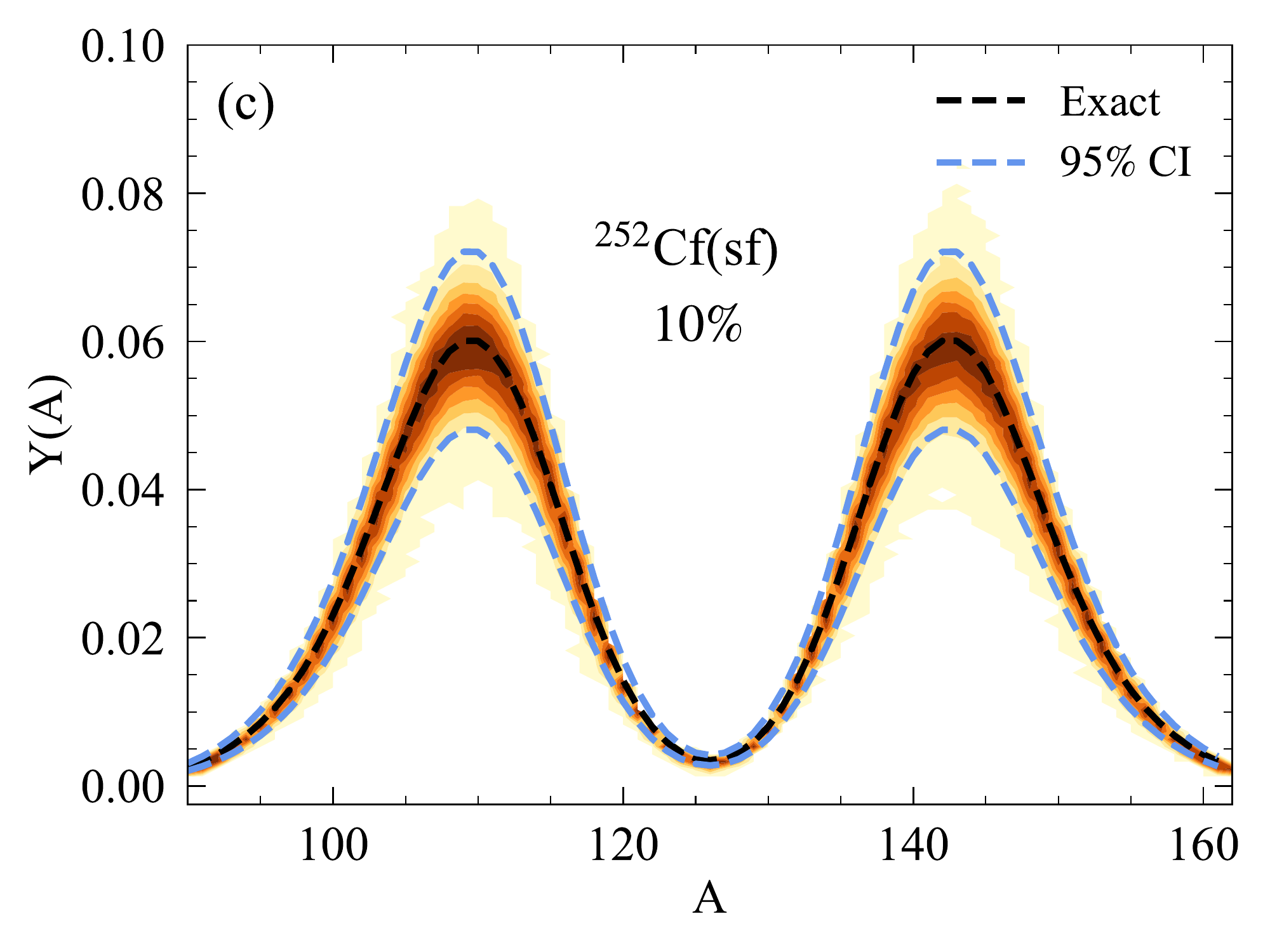} \\
\includegraphics[width=0.33\textwidth]{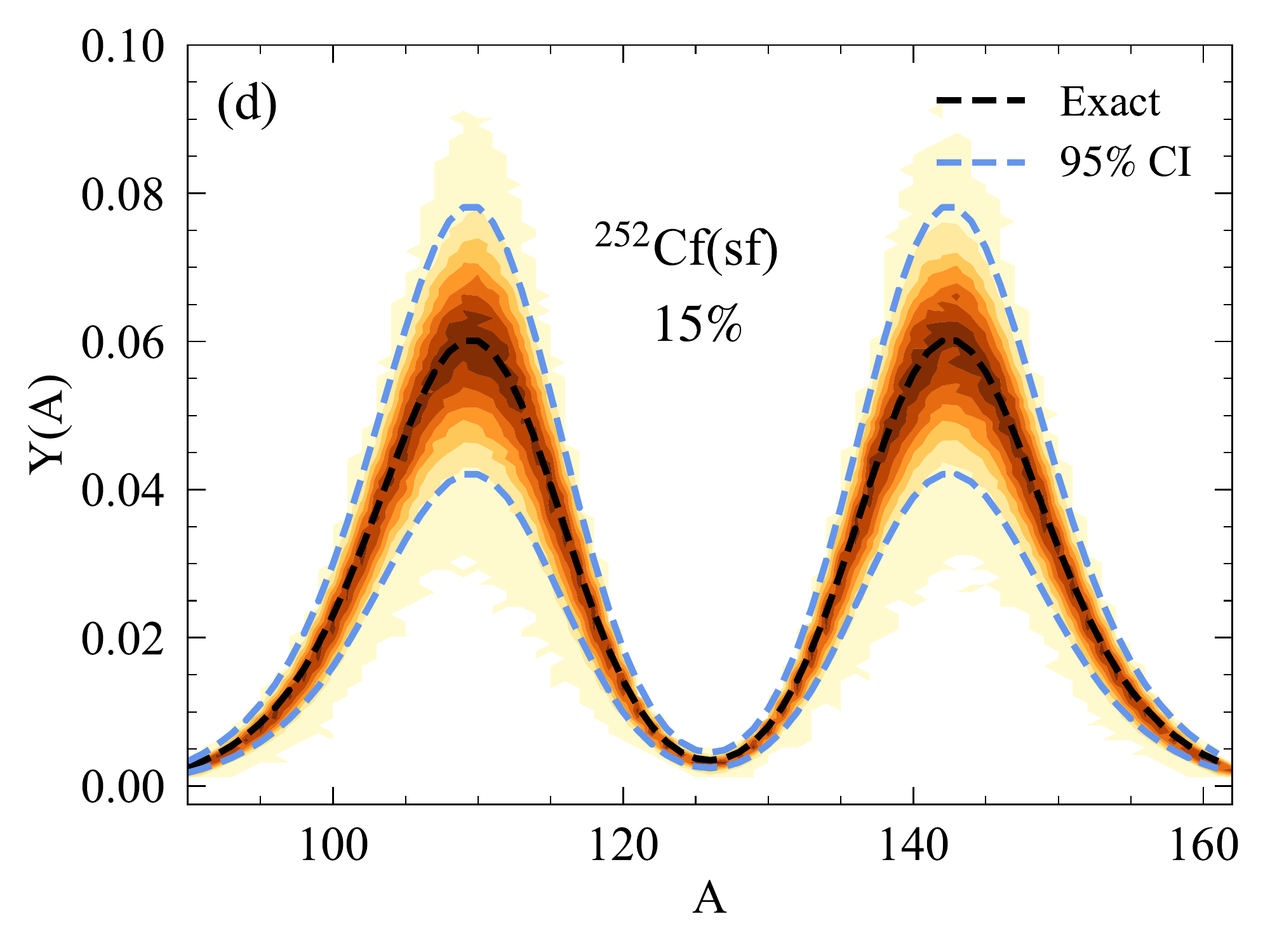} & \includegraphics[width=0.33\textwidth]{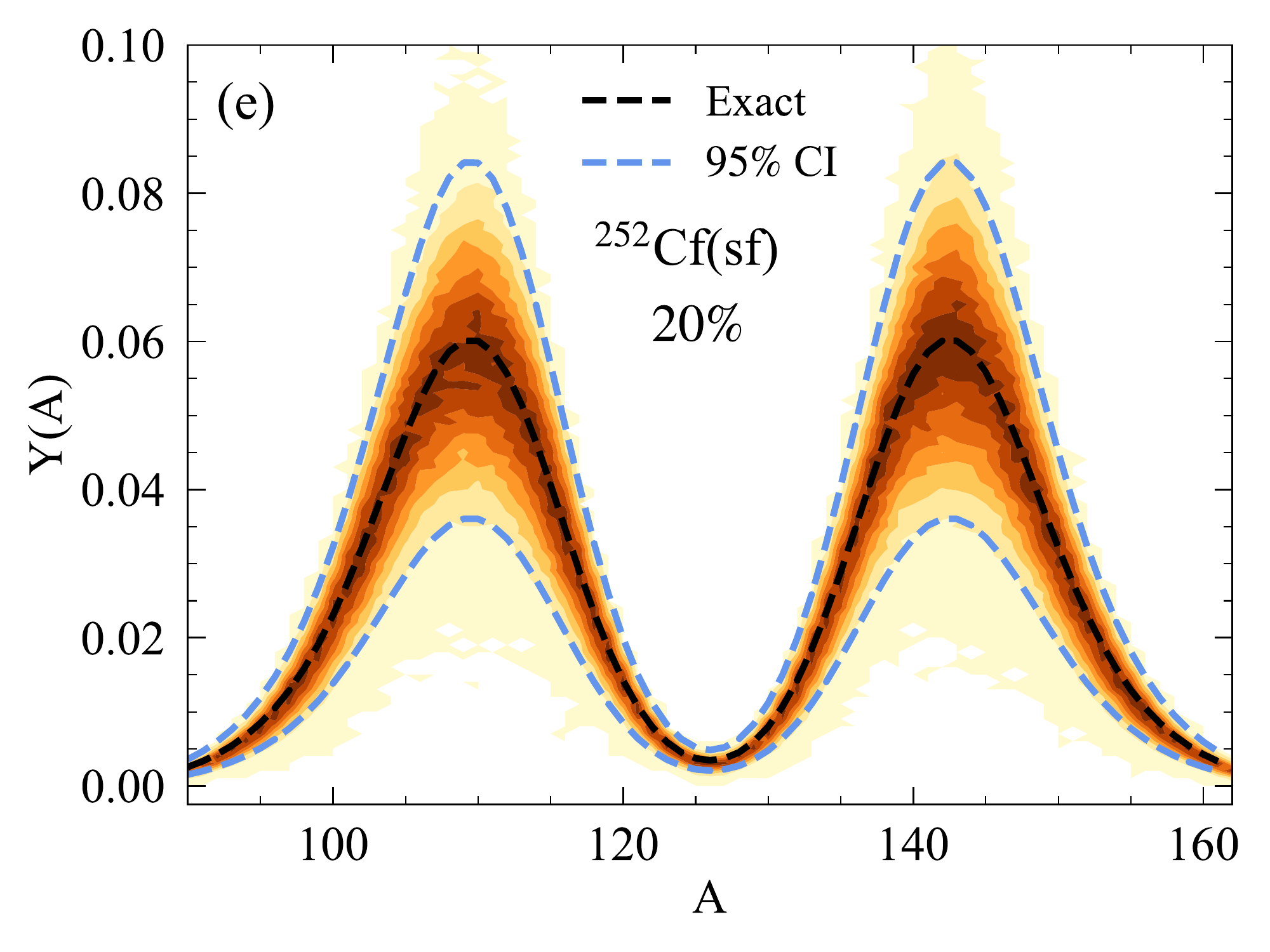} & \includegraphics[width=0.33\textwidth]{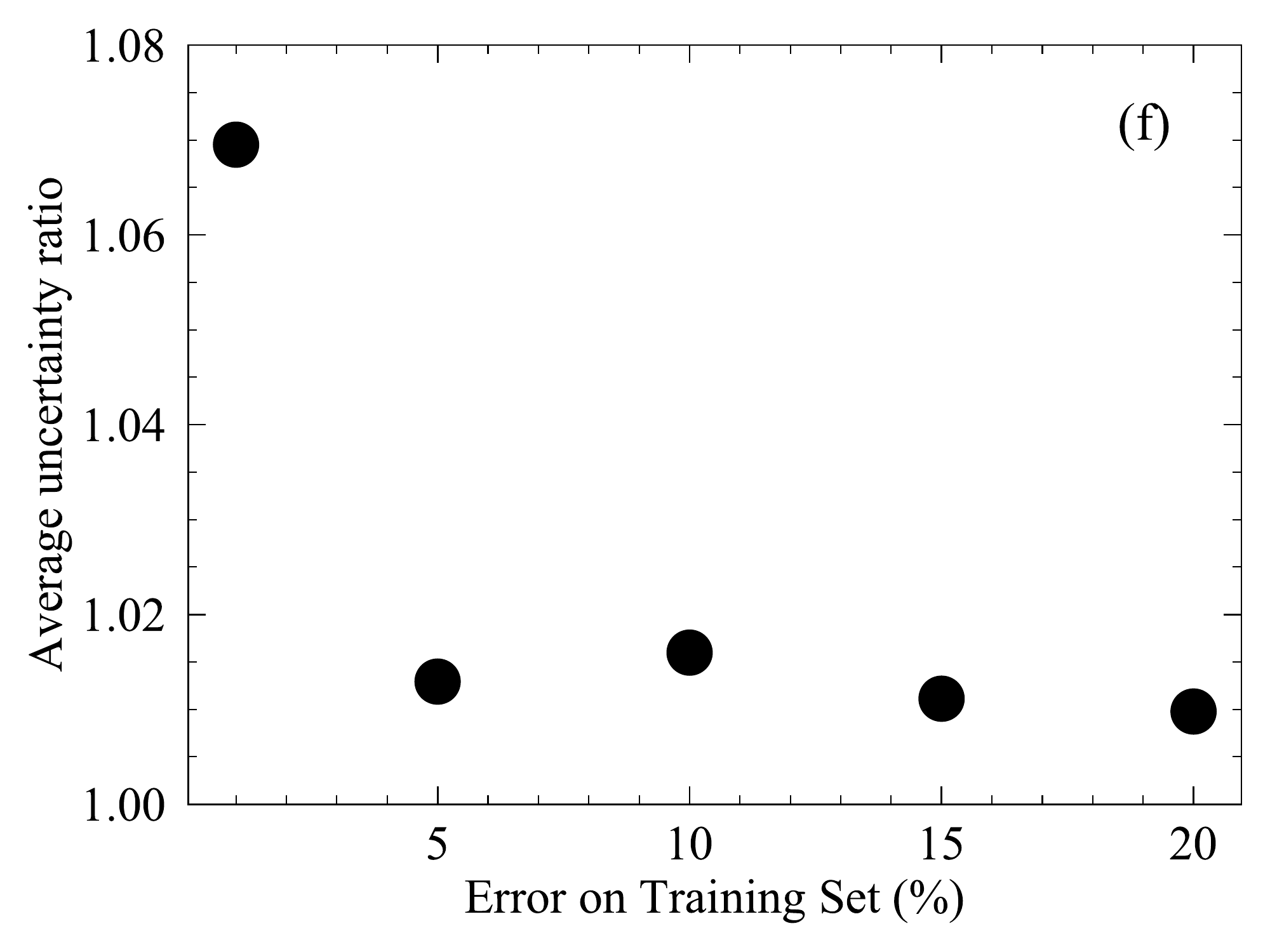} 
\end{tabular}
\caption{(color online) Density of the MDN prediction for the mass yields of $^{252}$Cf(sf) when various levels of uncertainty are considered, (a) 1\%, (b) 5\%, (c) 10\%, (d) 15\%, and (e) 20\% using 50 training points per mass.  Dashed black lines show the exact \texttt{CGMF} parametrization, and dashed blue lines give the 95\% confidence interval for the given level of uncertainty.  Panel (f), ratio of the average uncertainty, $\varepsilon_\mathrm{MDN}/\varepsilon_\mathrm{training}$ as defined in Eq. (\ref{eqn:averageUncertainty}), as a function of the percent uncertainty on the training set, for 50 samples per mass.}
\label{fig:Upropagation}
\end{figure}

Using 50 samples per mass in the training set, we can now study how the uncertainties propagate through the MDN when varying levels of uncertainty are taken on the training set.  Again, when we are just trying to reproduce the training set using the MDN, we would expect that the uncertainties on the predictions should exactly match the uncertainties on the training set, if the method is robust for propagating uncertainties.  In Fig. \ref{fig:Upropagation} we show a comparison between the training simulations and the predictions when these varying levels of uncertainty are assumed on the exact mass yields, (a) 1\% uncertainty, (b) 5\%, (c) 10\%, (d) 15\%, and (e) 20\% .  

For all levels of uncertainty, from 1\% to 20\%, the shape of the mass yields - as well as the spread in the training data set - is qualitatively reproduced by the MDN.  For a quantitative study, Fig. \ref{fig:Upropagation}(f) shows the ratio of the average uncertainty on the MDN prediction to the average uncertainty of the training set, $\varepsilon_{MDN}/\varepsilon_\mathrm{training}$, as a function of the percent uncertainty on the training set.  As seen in the plot, $\varepsilon_\mathrm{MDN}$ is 7\% higher than $\varepsilon_\mathrm{training}$ when 1\% uncertainties are assumed.  As the uncertainty increases, the discrepancy drops below 2\%.  This discrepancy can be further reduced by increasing the yield cut-off for the lowest mass yields (currently set near $10^{-3}$) - again indicating that the scaling in the training set is causing these difficulties.

\subsection{Reproducing experimental data}
\label{sec:experimentalData}

\begin{figure}
\centering
\begin{tabular}{cc}
\includegraphics[width=0.5\textwidth]{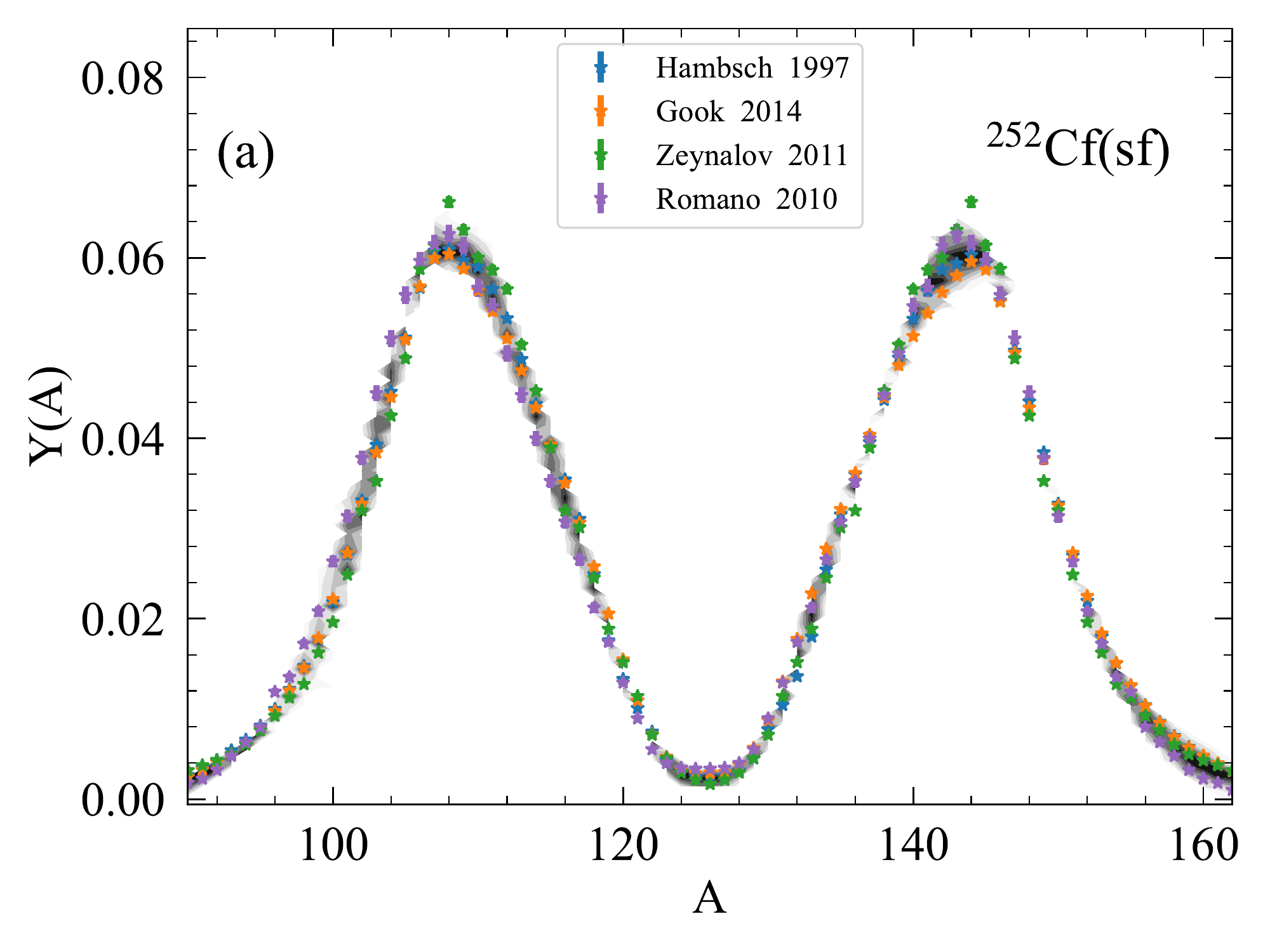} \includegraphics[width=0.5\textwidth]{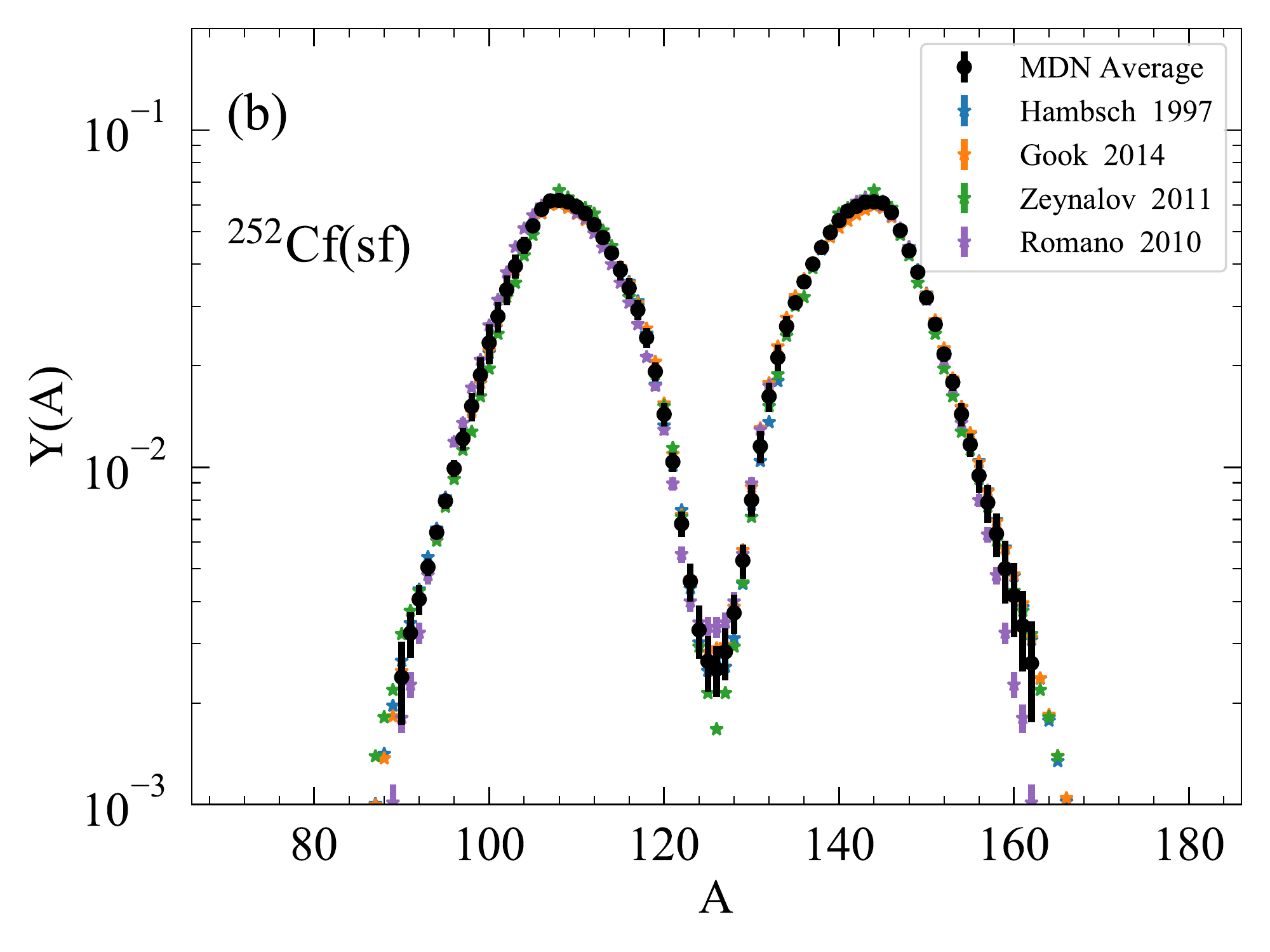} 
\end{tabular}
\caption{(color online) (a) Comparison between the training set and the density of the resulting MDN calculation for the mass yields for $^{252}$Cf(sf).  (b) Comparison between the average of the MDN calculation and the experimental data sets used to construct the training set.}
\label{fig:252CfExp}
\end{figure}

Experimentally, $^{252}$Cf(sf) mass yields have been very well measured by several groups.  To study the MDN with a training set constructed from experimental data, we chose four measurements that reported data for the mass yields of the light and heavy fragments along with experimental errors.  As with the three-Gaussian parameterization, to take the uncertainties into account, we construct the training set by sampling the experimental yields from a Gaussian distribution within their reported uncertainties.  At this point, we assume no correlation between the masses within a single data set and no correlation between each data set.  The colored stars in Fig. \ref{fig:252CfExp}(a) show the four experimental data sets.  Here, we can see some inconsistencies between the four data sets, primarily when the yields are small.  Although the differences among the data sets are minimal, it is worth noting that the experimental error bars, which are mostly statistical, do not necessarily overlap in these regions.  These discrepancies then provide an opportunity to investigate how the MDN handles discrepant data sets.

To construct the training set, 100 samples are pulled from a Gaussian distribution defined at each experimental value.  The yields at each mass are sampled independently from one another, regardless of whether they are within the same experiment or different experiments.  Once again, we have a choice of how many Gaussian mixtures to use.  Because there are four data sets that comprise the training set, we use four Gaussian mixtures, to allow for the possibility that each data point will be described by a separate Gaussian mixture.  

In Fig. \ref{fig:252CfExp}(a) we also show a comparison of the MDN prediction and the four data sets.  Here, we begin to see that the MDN does not simply reproduce the spread in the training set for each mass but constructs areas of higher density based on which regions are more likely.  For each $A$ value, we can calculate the average of the MDN and the corresponding standard deviation.  These are shown in panel (b) in comparison to the four sets of experimental data.  There is no mass range where the MDN average is an outlier compared to the experimental values.  

\begin{figure}
\centering
\includegraphics[width=0.5\textwidth]{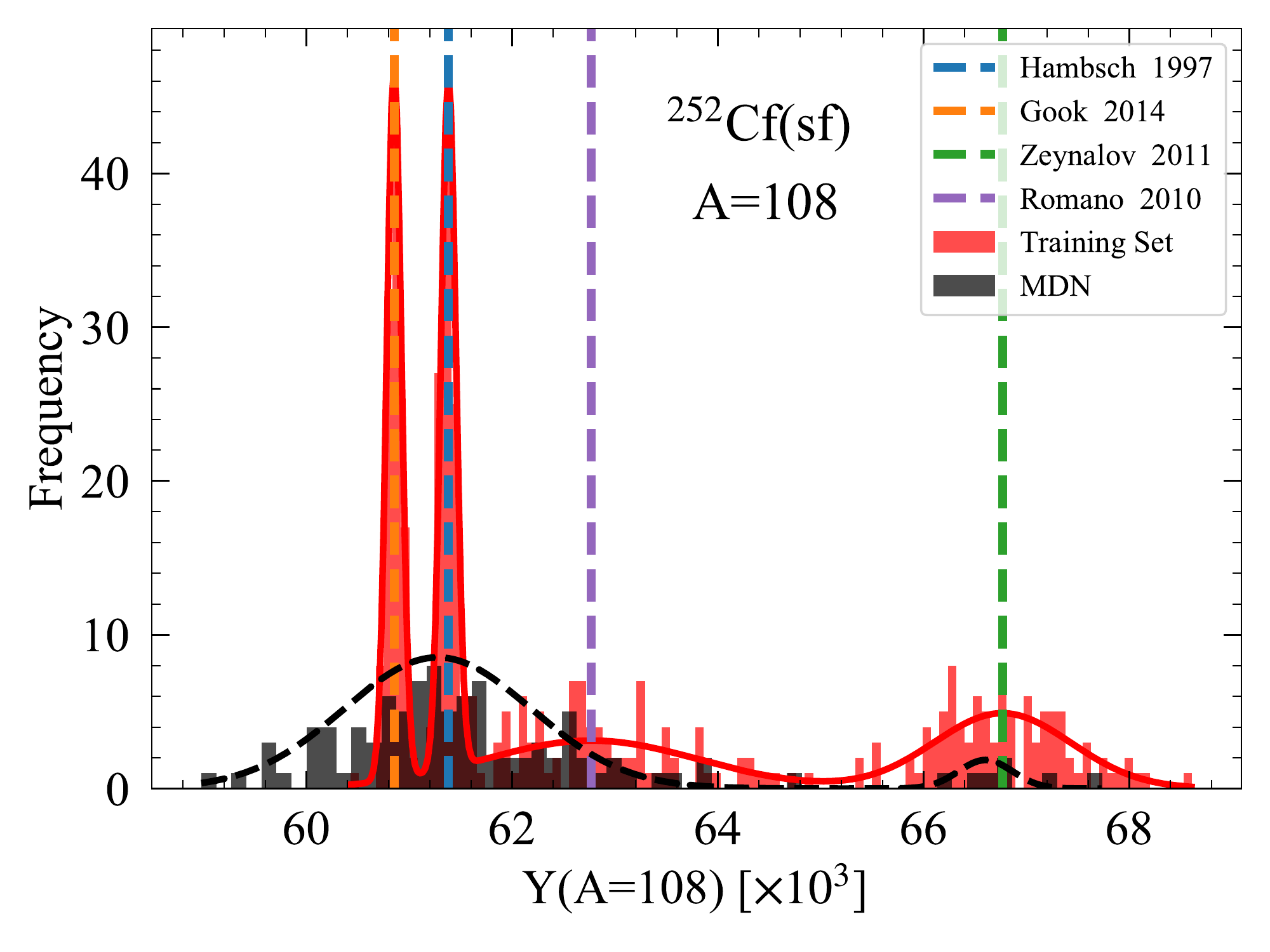}
\caption{(color online) Comparison of the distribution of training data and the posterior distribution of the resulting MDN predictions for a single fragment mass, $A=108$.}
\label{fig:252CfA108}
\end{figure}

To look more closely at how the MDN handles the discrepancies between the data sets - as encoded in the training set - in Fig. \ref{fig:252CfA108}, we plot the posterior distribution of the MDN in black compared to the training set, in red, for one mass value $A=108$ which is in the light peak of the mass yield for $^{252}$Cf(sf).  The mean values for each data set are shown as dashed vertical lines.  In the training set, we see a clear distinction between the four experimental data sets, including very little overlap between the data set of Zeynalov (mean indicated by the green vertical line) and the other three data sets.  We find that the posterior distribution of the MDN does not just reproduce the distribution of the training set.  In addition, the mean of the MDN prediction, 0.0618, is not just simply the mean of the training data, 0.0629.  The MDN is inferring the most likely distribution given the experimental data from the flexibility of the four Gaussian-mixture that is included.  

\subsection{Testing physical constraints on the mass yields}

Although there is a good qualitative comparison between the training set and MDN prediction, there are two main quantitative constraints that the mass yields should uphold.  The first is that they should be normalized, as they represent a probability distribution, and the second is that they should be symmetric (for spontaneous fission and neutron-induced fission at low incident energies).  In this work, all of the mass yields have been normalized to 2, and for $^{252}$Cf(sf), the yields should be symmetric around $A=A_c/2=126$.

\begin{figure}
\centering
\begin{tabular}{cc}
\includegraphics[width=0.5\textwidth]{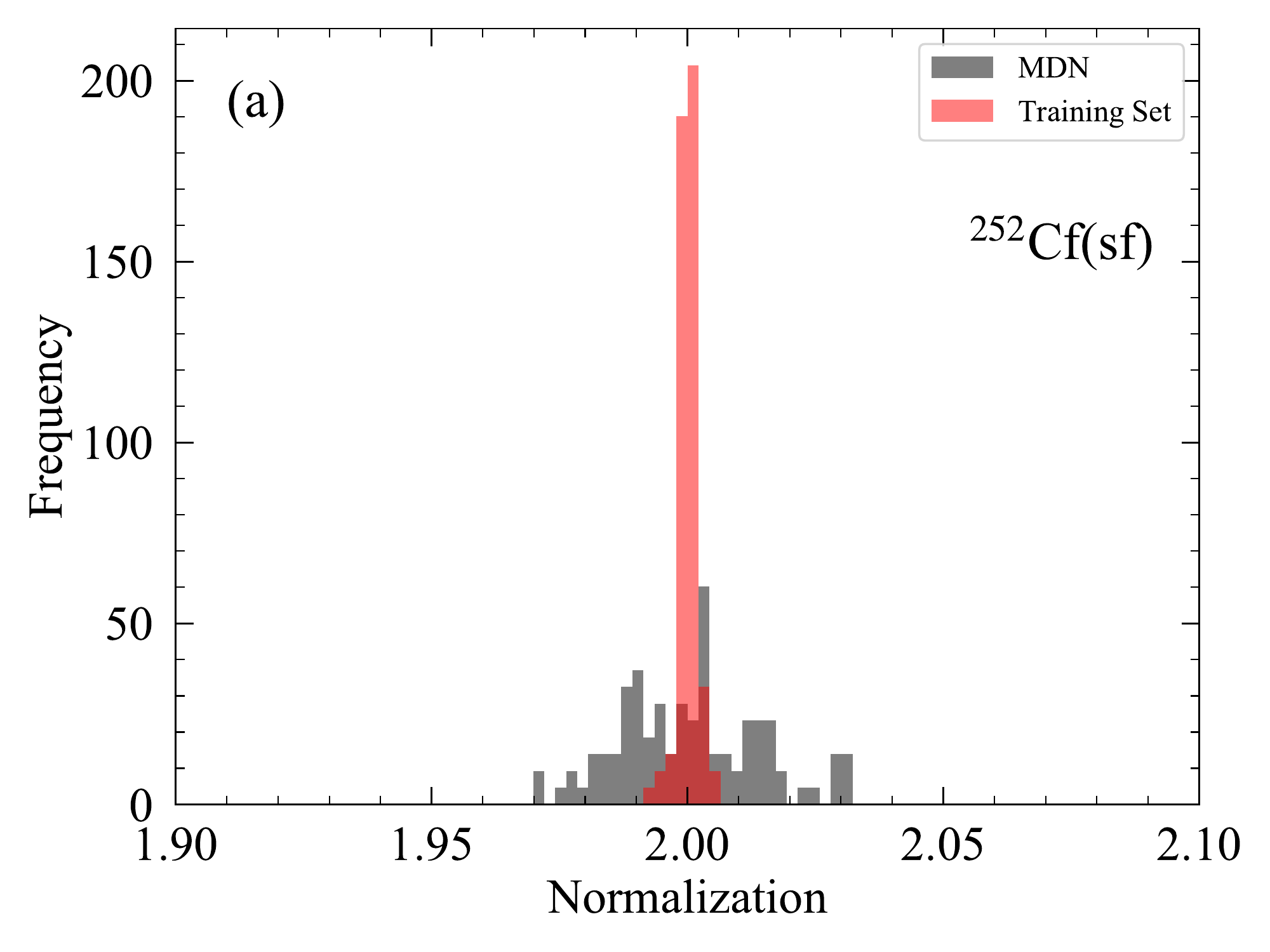} & \includegraphics[width=0.5\textwidth]{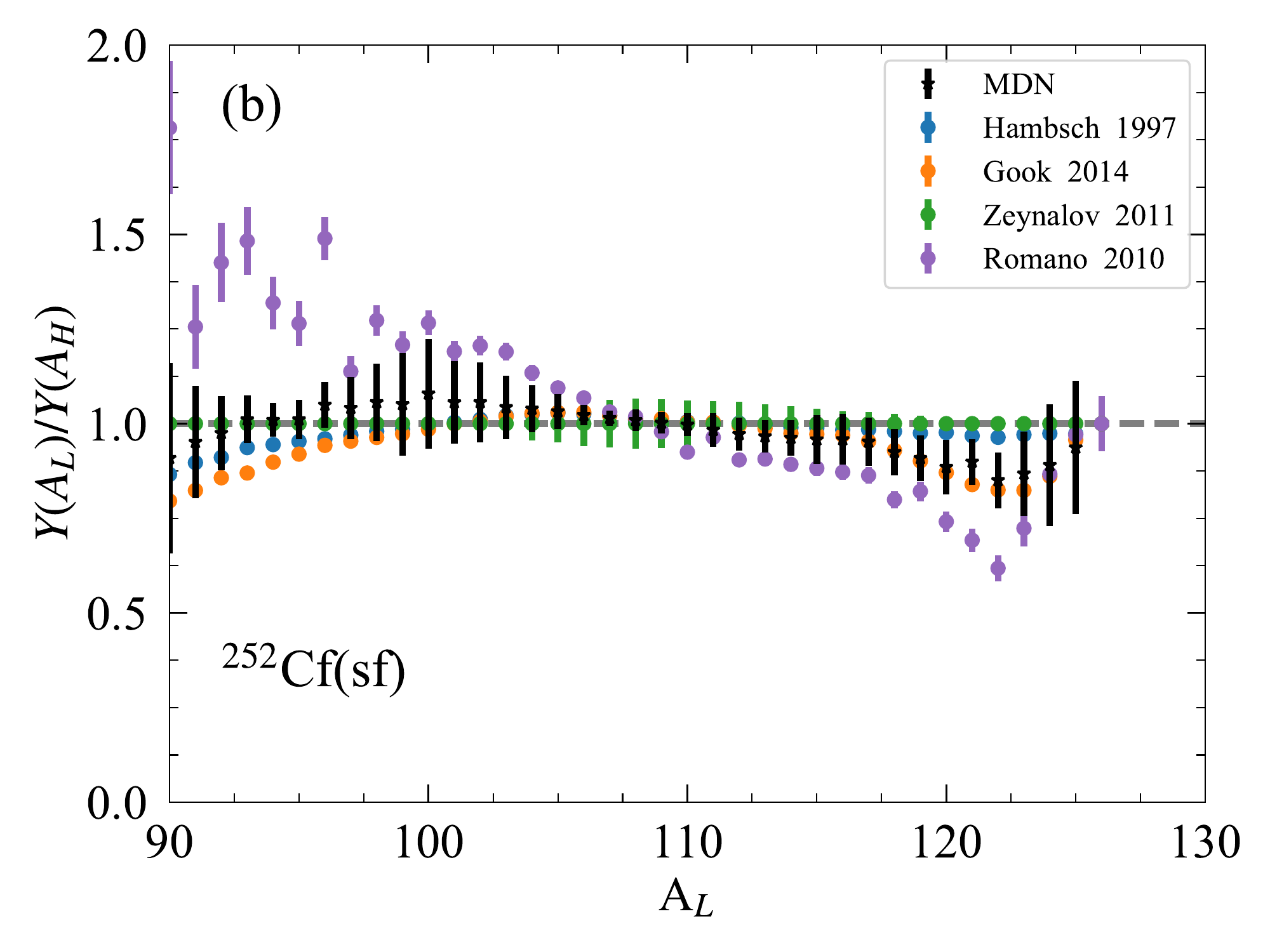}
\end{tabular}
\caption{(color online) (a) Comparison of the distributions of normalization of each sample from the MDN, grey, to the normalization of the training set, red.  (b) Ratio of the yields of the light to the heavy fragments (symmetry test) for the MDN (black stars) compared to the experimental data sets (colored circles).}
\label{fig:252CfPhysics}
\end{figure}

In Fig. \ref{fig:252CfPhysics}(a), we show the distribution of the normalizations of the training set, red, compared to the MDN results, grey.  In both cases, because the yield at each $A$ value is sampled independently from the rest of the distribution, there is an ambiguity in what is meant by the normalization of a single yield distribution.  The yield at each mass in the training set has been sampled the same number of times, so we select one value of the yield for each mass and calculate the normalization based on that distribution.  The normalization of the training set is within 0.5\% of the expected value of 2.0, while the normalization of the resulting MDN samples is within about 1.5\% of the expected normalization.

The second condition that the mass yields must fulfill is that they are symmetric around $A=A_c/2$, with $A=126$ for $^{252}$Cf(sf).  In Fig. \ref{fig:252CfPhysics}(b), we show the mass yield for the light fragment divided by the mass yield of the corresponding heavy fragment, $Y(A_L)/Y(A_H)$ where $A_H = A_c - A_L$.  For symmetric mass yields, this ratio should be 1.0, but we can see in Fig. \ref{fig:252CfPhysics} (b) that this ratio is not held for all masses, neither for the experimental values nor the MDN predictions.  The experimental data deviate further from 1.0 in the tails and the symmetric region of the mass distributions, where the yields are small and much harder to measure (and the experimental uncertainties are much larger).  The ratio of the MDN prediction is bounded by experimental ratios, most closely following the ratios of the Hambsch \cite{Hambsch1997} and G{\"o}{\"o}k \cite{Gook2014}.  This is reassuring as these two data sets have the lowest experimental uncertainty, and we would expect the MDN algorithm to be driven by the most accurate data.    

These two conditions could both be easily ensured by only training on and predicting one half of the mass yields (either the heavy or the light fragment).  Even so, it is promising that some of the physics is already incorporated into the MDN prediction without enforcing it directly.  We can also build more physics into the training set by sampling the yields from a covariance matrix instead of drawing each mass independently.  In the next section, we explore a first step to building these physical constraints directly into the training set.

\subsection{Physics-driven feature engineering of the training dataset}
\label{sec:constraints}

In the previous sections, each $Y(A)$ value was sampled independently from every other mass in the distribution.  However, we know that there are correlations between the masses through the normalization and symmetry conditions, as well as experimental covariances (which are rarely reported).  In this section, we take the first step to including these physical constraints on the training set by building in the symmetry between the mass yields of the heavy and light fragments and by building in the normalization condition for each sample of the training set.  To ensure that the mass yields are symmetric, we sample the light fragment mass yield only and constrain the complementary heavy fragment to have the same yield.  Then the sampled mass distribution is rescaled such that $\sum \limits _i Y(A_i) = 2$.

\begin{figure}
\centering
\begin{tabular}{cc}
\includegraphics[width=0.5\textwidth]{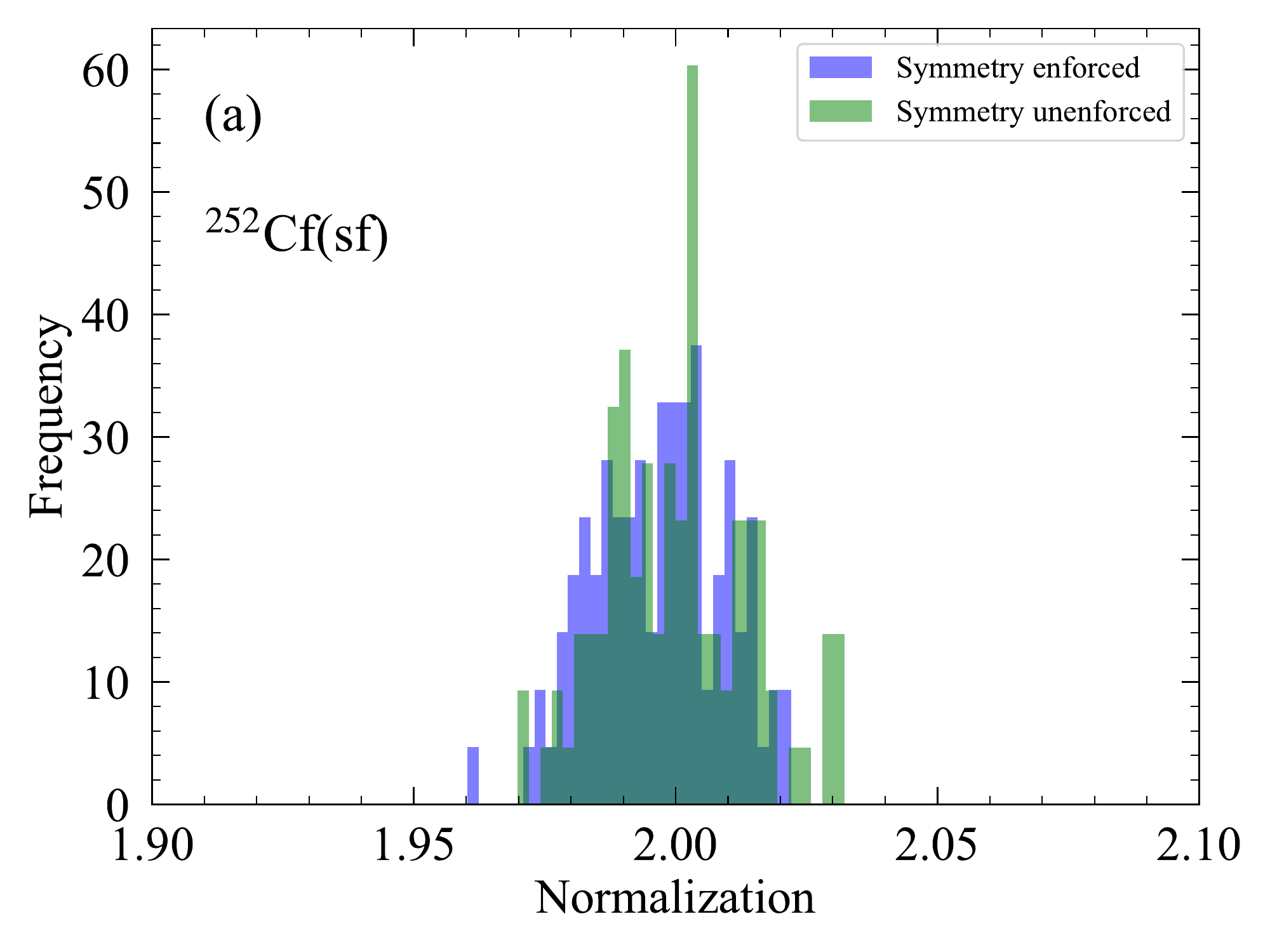} & \includegraphics[width=0.5\textwidth]{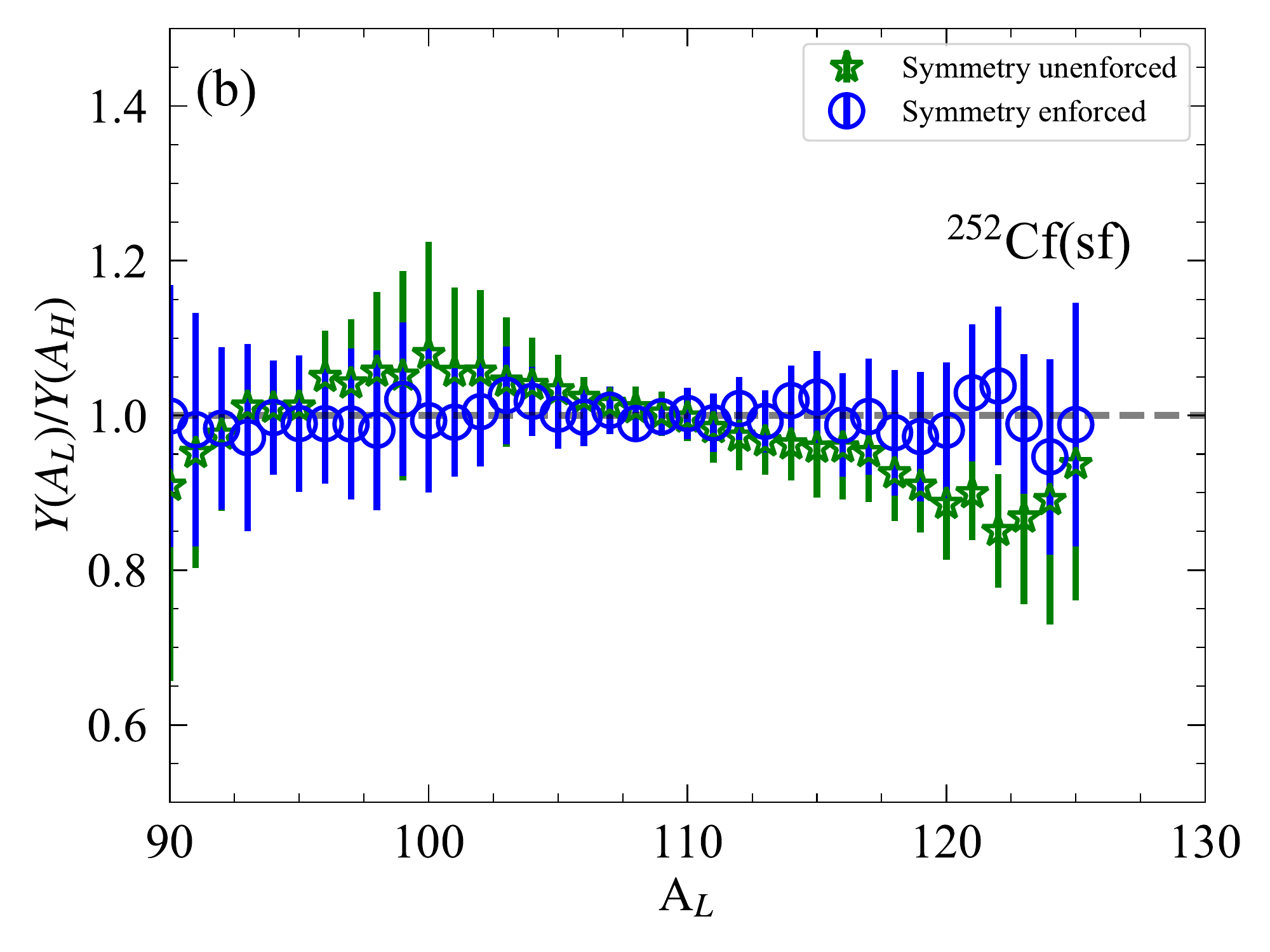} \\
\end{tabular}
\caption{(color online) Comparison of (a) the normalization of each sample from the MDN and (b) the symmetry of the mass yields, when the training set was sampled independently, green, and when the normalization and symmetry constraints were built into the training set, blue.}
\label{fig:252CfConstraints}
\end{figure}

To show the effect on the MDN prediction, we first explore how the normalization and symmetry of the MDN predictions change when these constraints are included in the training set.  In Fig. \ref{fig:252CfConstraints}(a), the normalization of the MDN predictions trained on the unconstrained mass yields (described in the previous subsection) and trained on the constrained mass yields are compared, blue and green distributions respectively.  Even though the normalization was built in exactly into the training set, the normalization of the MDN prediction was not any more constrained; in both cases, the normalization was $2.00 \pm 0.01$.  In Fig. \ref{fig:252CfConstraints}(b), we compare the ratios of the mean mass yields for the light fragment and the complementary heavy fragment; blue circles show the ratio for the constrained MDN prediction, and green stars show the unconstrained MDN prediction.  Here, there is an improvement on the MDN prediction when the training set has the symmetry constraint included directly into the training set.  The mean of the ratio goes from 0.98 to 1.00 when the symmetry is included in the training set, and the deviation from $Y(A_L)/Y(A_H) = 1$ decreases from 0.06 to 0.02 when the symmetry is included.


As a second step, we could construct a covariance matrix from the known experimental uncertainties and correlations and sample the training set from the covariance matrix directly.  This should impose further constraints on the training set which could then be reflected in the MDN predictions.  Moving forward, this will be explored, particularly in the context of comparing the MDN to other machine learning and optimization methods.

\section{Interpolation and extrapolation from the training set}
\label{sec:interp}

Being able to reproduce the training set is an important part of validating any machine learning algorithm, but ultimately to be useful, we also have to be able to make reasonable predictions outside of our training set.  To test the reliability of the MDN for interpolating between and extrapolating beyond samples in the training set, we now turn to the energy dependence of $^{235}$U(n,f) mass yields.  Instead of having a single input feature, mapping $\{A\}$ to $\{Y(A)\}$, we now have two input features, mapping $\{A,E_\mathrm{inc}\}$ to $\{Y(A|E_\mathrm{inc})\}$.  Our construction of the MDN in \texttt{PyTorch} allows for an arbitrary number of input features, so using two features instead of one does not take any updating.

\begin{figure}
\centering
\begin{tabular}{cc}
\includegraphics[width=0.5\textwidth]{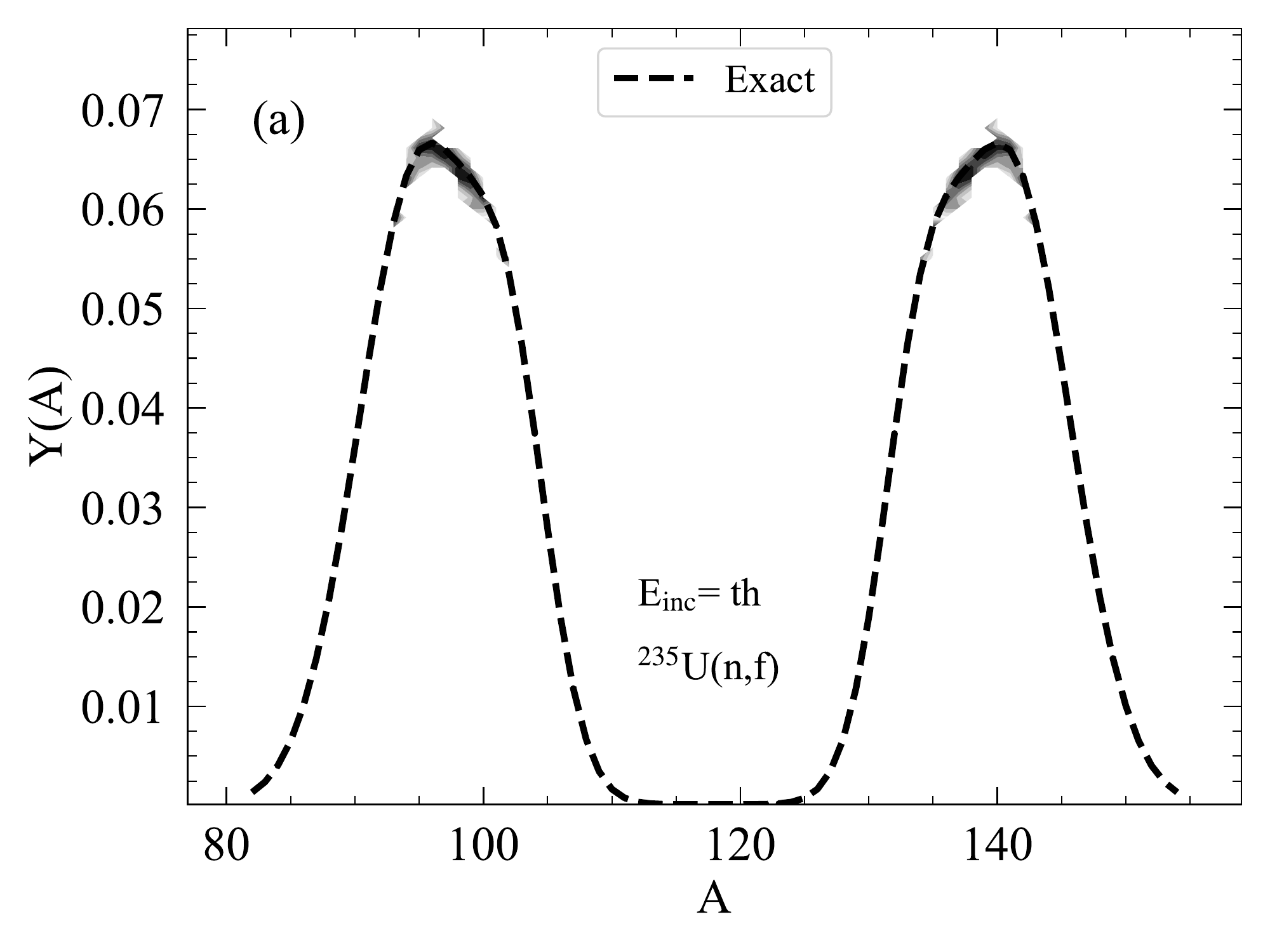} & \includegraphics[width=0.5\textwidth]{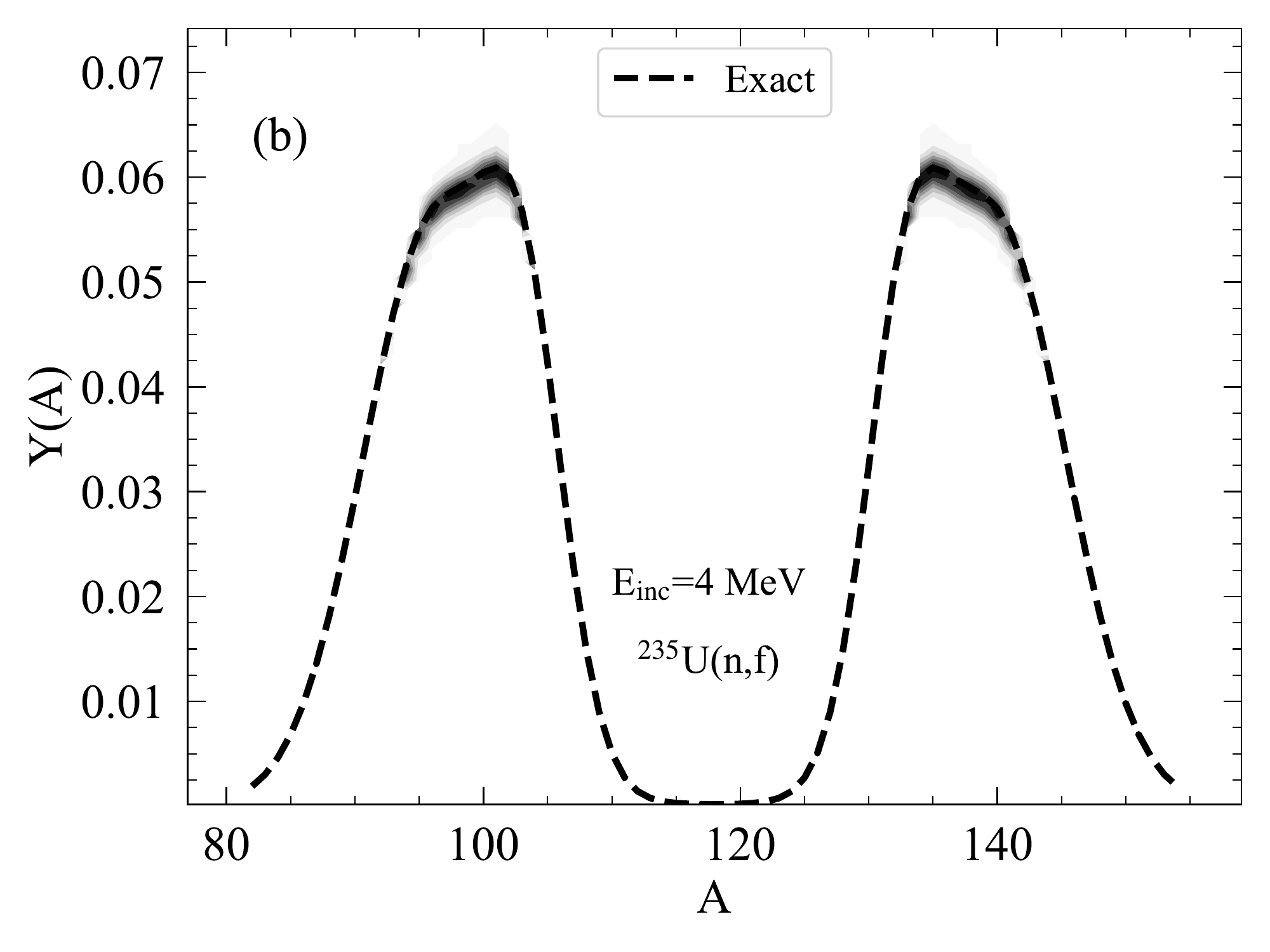} \\
\includegraphics[width=0.5\textwidth]{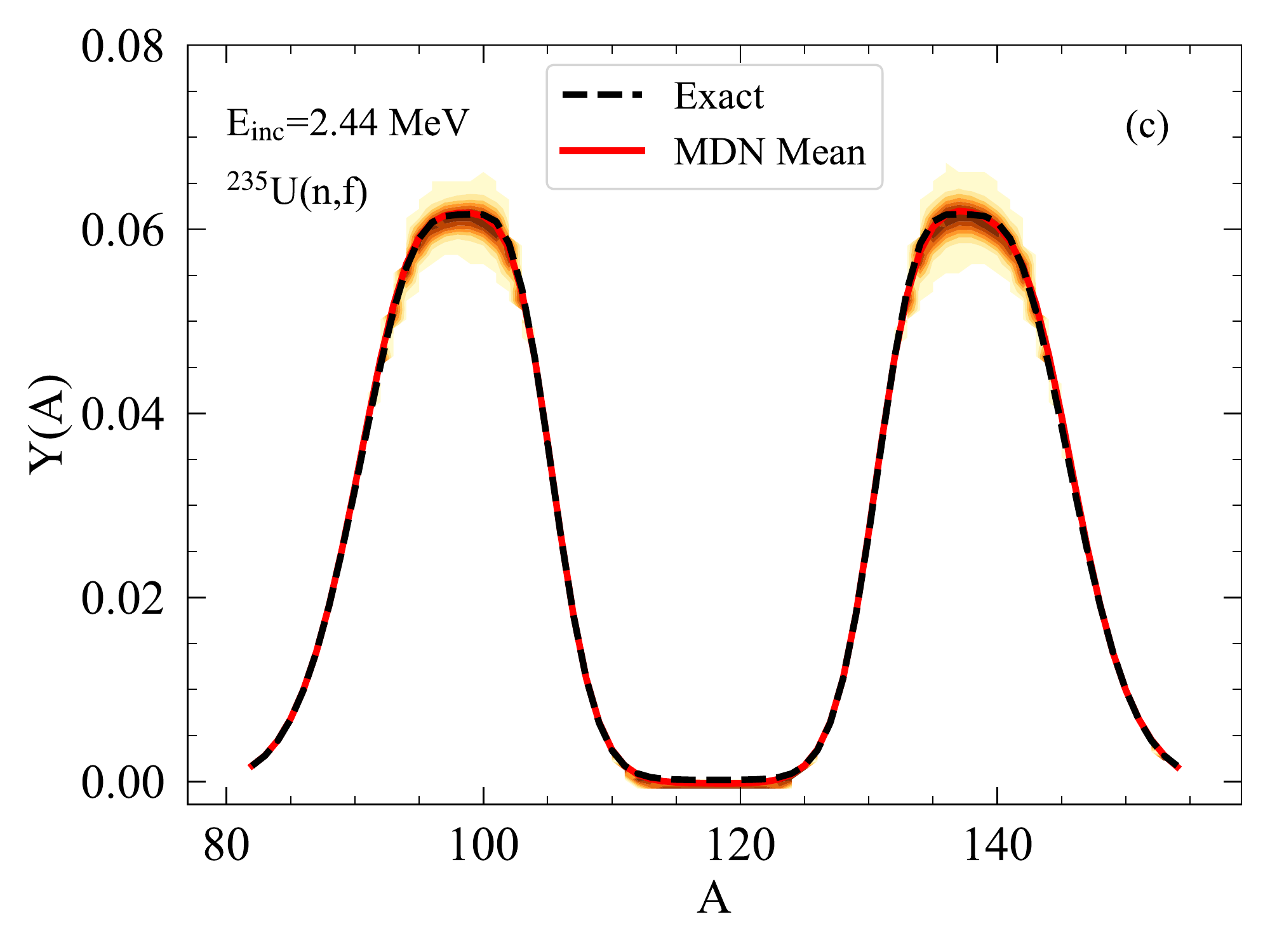} & \includegraphics[width=0.5\textwidth]{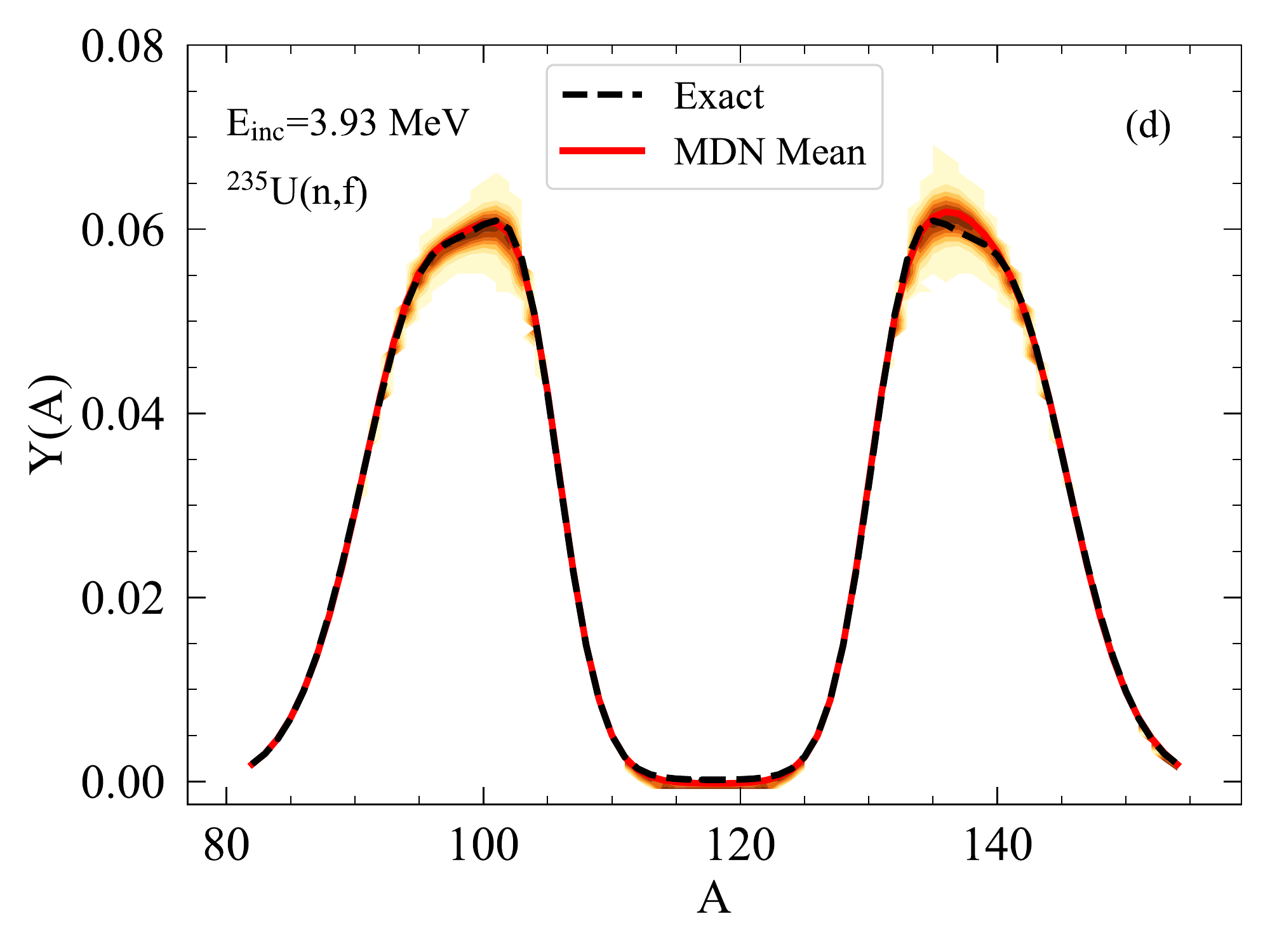} \\
\includegraphics[width=0.5\textwidth]{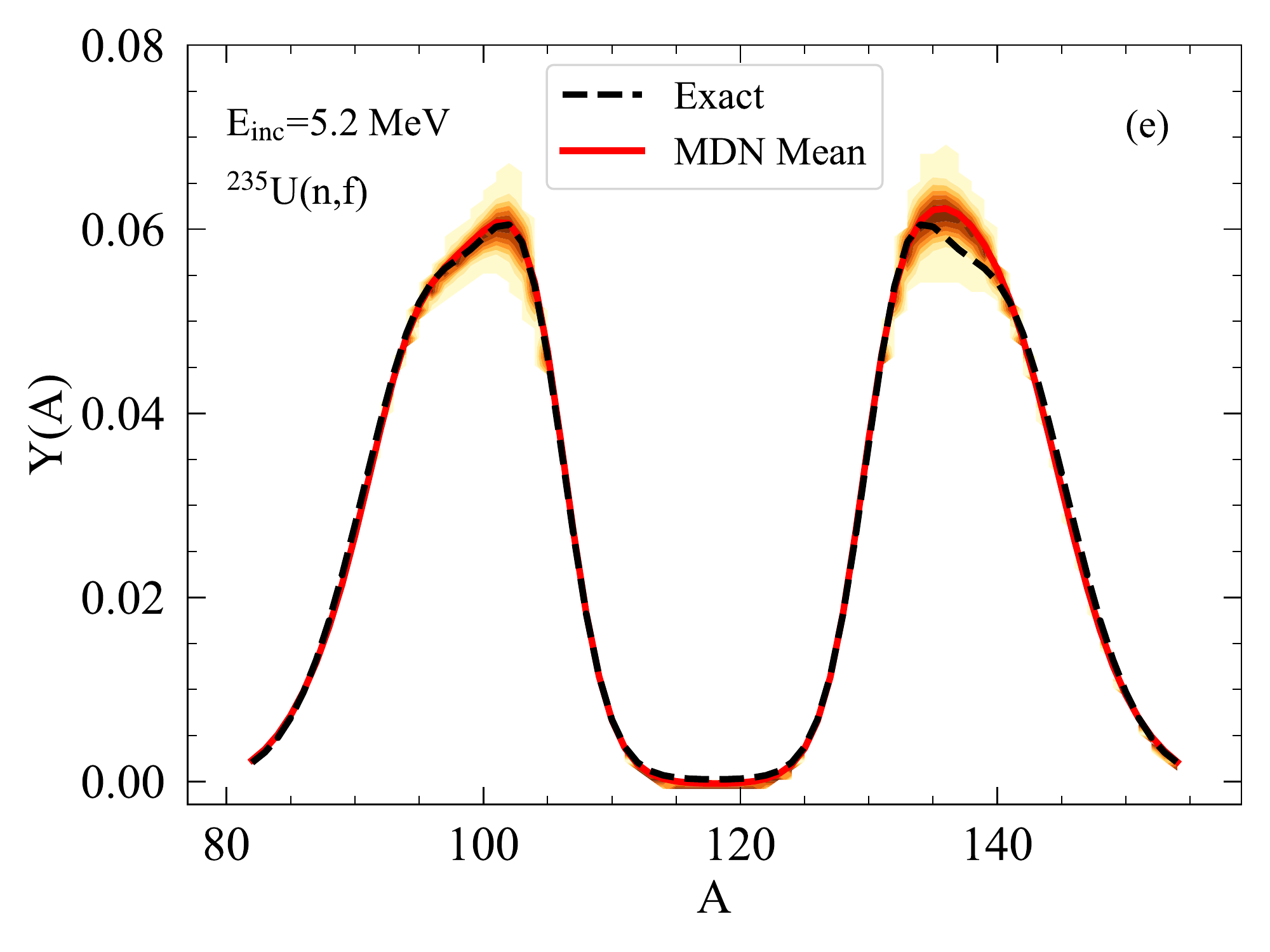} & \includegraphics[width=0.5\textwidth]{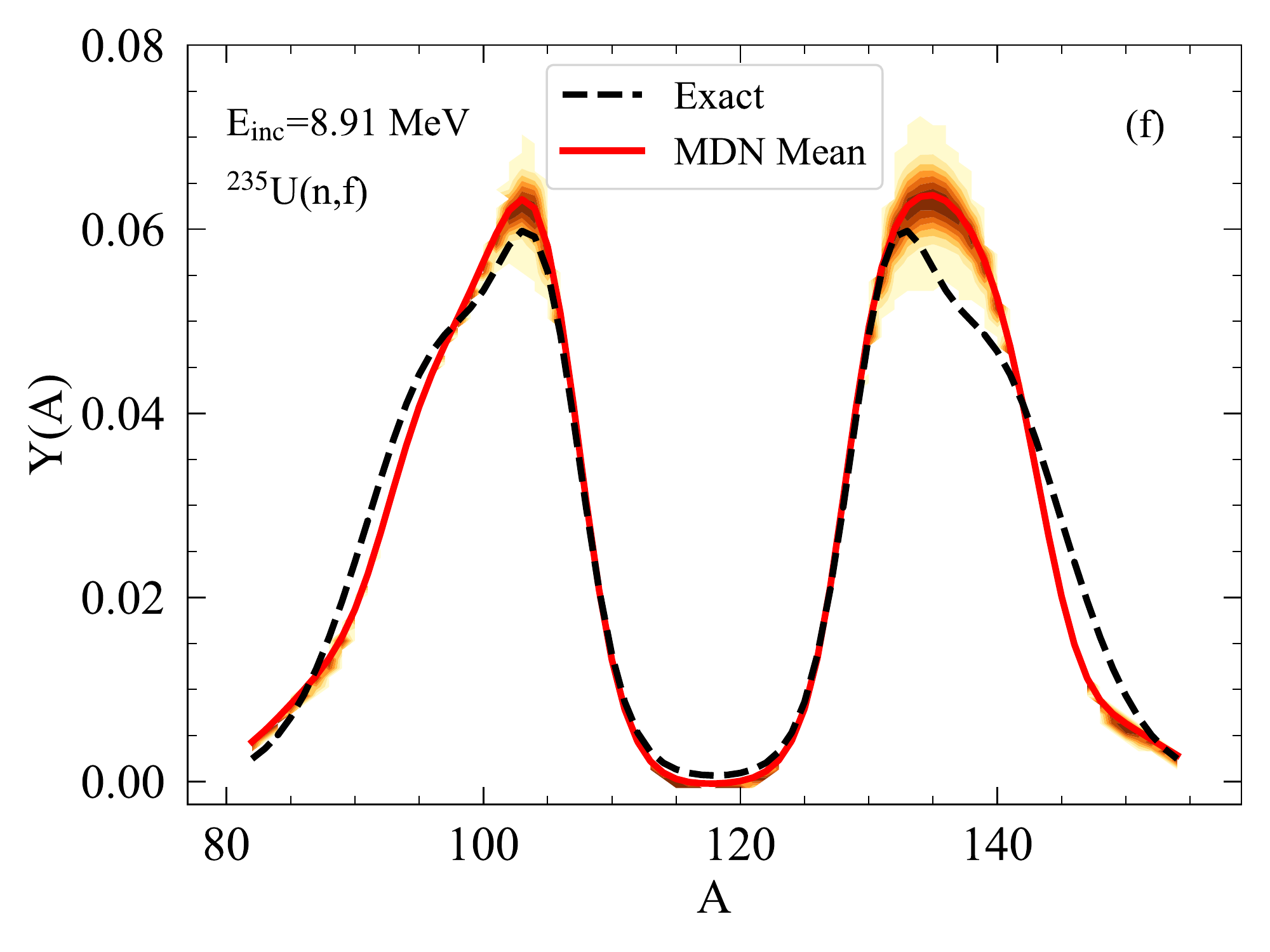}
\end{tabular}
\caption{(color online) Density distribution of the training set compared to the exact $Y(A)$ parameterization of $^{235}$U(n,f) for (a) E$_\mathrm{inc}$= thermal and (b) $E_\mathrm{inc}$=4 MeV.  Resulting interpolated predictions from the MDN compared to the exact $Y(A)$ parameterization for (c) $E_\mathrm{inc}$=2.44 MeV and (d) $E_\mathrm{inc}$=3.93 MeV.  Extrapolations beyond the training set for the MDN predictions (orange density and red solid lines to indicate the average value of the distribution) compared to the exact $Y(A)$ parameterization - without multi-chance fission taken into account for (e) $E_\mathrm{inc}$=5.20 MeV and (f) $E_\mathrm{inc}$=8.91 MeV.}
\label{fig:235U}
\end{figure}

We build a training set of mass yields for $^{235}$U(n,f) for an incident neutron energy grid between thermal and 4 MeV in steps of 1 MeV, below second-chance fission, where a neutron is emitted from the compound nucleus before it fissions.  For each incident energy, the mass yields were again constructed from the three-Gaussian parametrization of Eq. (\ref{eqn:threeGaussian}) and values were sampled independently within 2\% of the nominal value, constrained to be symmetric about $A_c/2$.  Two examples from the training set are shown in Fig. \ref{fig:235U} for (a) E$_\mathrm{inc}$=2.53$\times 10^{-8}$ MeV (thermal) and (b) E$_\mathrm{inc}$=4 MeV.  As with the $^{252}$Cf(sf) training sets, we removed masses for which the yield was less than $10^{-3}$; for the training set, this means that the yields for some of the symmetric mass region were removed - not the case for $^{252}$Cf(sf) mass yields.  If this region is kept, the quality of the training is significantly reduced, again indicating that a different scaling is needed for the mass yields.

After training the MDN on this set, we can see how well the mass yields for non-integer incident energies are predicted.  We randomly sample values for the energy in the interval from thermal to 10 MeV.  These predictions are interpolations within the training set and extrapolations for incident energies above those included in the training set.  In addition, we also predict the yields for the symmetric mass region that was removed from the training set.  In Fig. \ref{fig:235U}, we show two examples from the testing set for interpolated energies, at (c) E$_\mathrm{inc}$=2.44 MeV, roughly equally spaced between two energies in the training set, and (d) E$_\mathrm{inc}$=3.93, nearly identical to one of the energies in the training set.  Finally, the last set of plots in Fig. \ref{fig:235U} shows two examples from the testing set at energies that are extrapolated beyond the training set, (e) $E_\mathrm{inc}$=5.20 MeV and (f) $E_\mathrm{inc}$=8.91 MeV.  

\begin{figure}
\centering
\begin{tabular}{cc}
\includegraphics[width=0.5\textwidth]{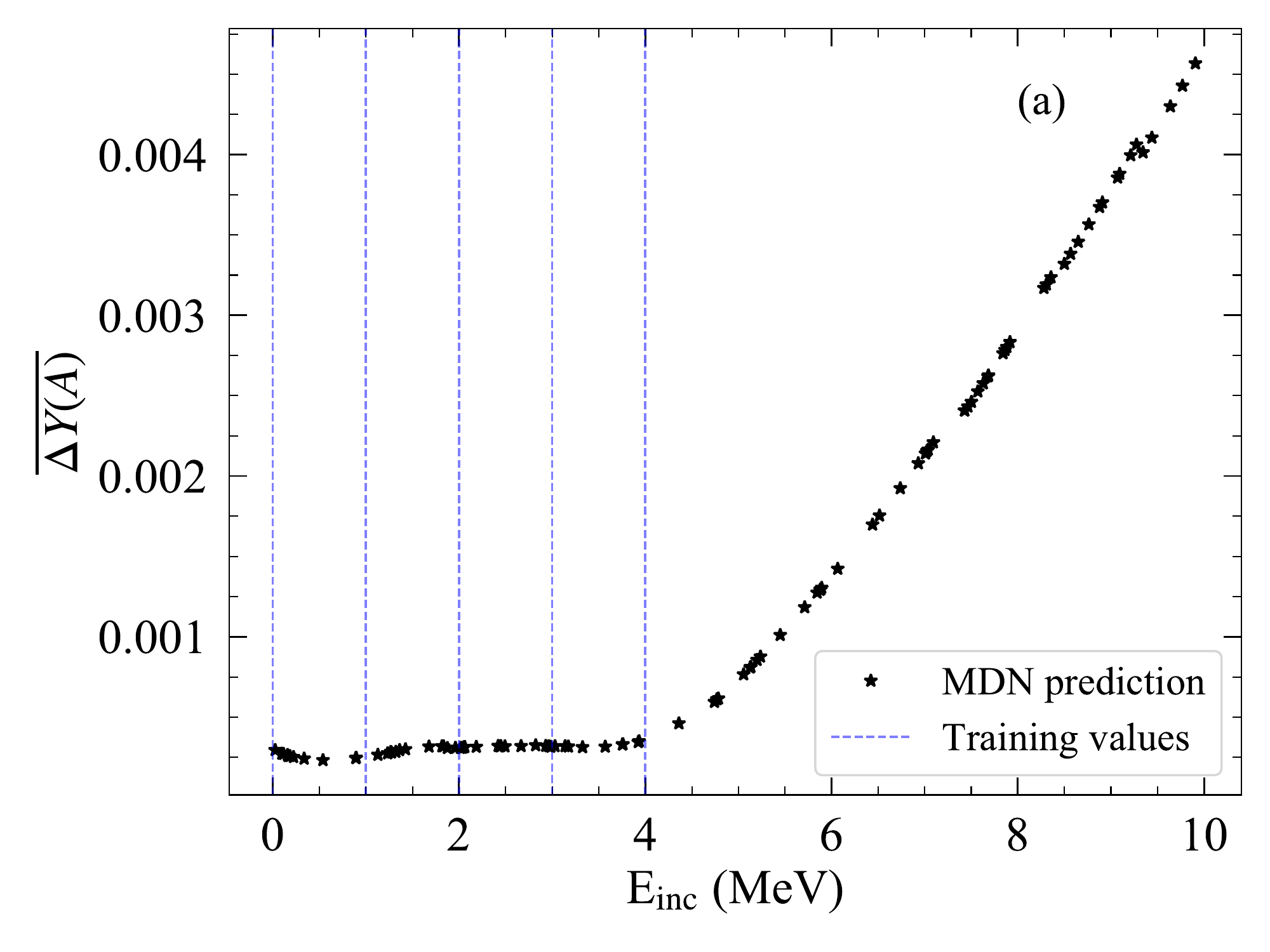} & \includegraphics[width=0.5\textwidth]{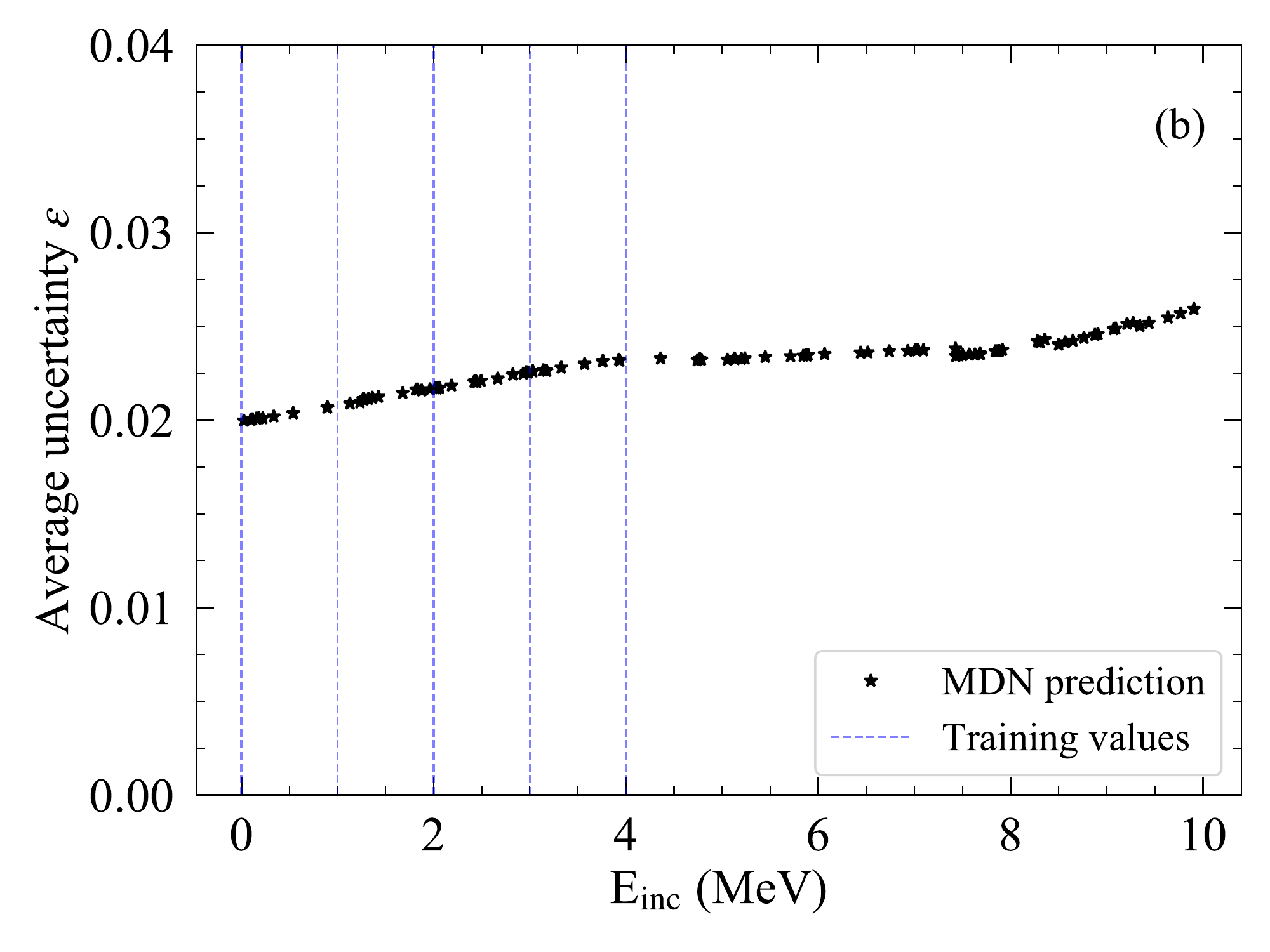} 
\end{tabular}
\caption{(color online) (a) Average difference between the training sets and the MDN predictions, Eq. (\ref{eqn:avgDeltaY}), as a function of neutron incident energy for the MDN predictions (black stars).  Incident energies included in the training set are indicated by the blue vertical lines.  (b) Average uncertainty, $\varepsilon$ Eq. (\ref{eqn:averageUncertainty}), on the MDN predictions (black stars); again the incident energies included in the training set and indicated by dashed vertical blue lines.}
\label{fig:235UStatistics}
\end{figure}

For predictions where the incident neutron energy falls between those energies included in the training set, there is a very good agreement between the mean values predicted by the MDN and the exact energy-dependent $Y(A)$ parameterization, although we begin to see discrepancies at energies near 4 MeV.  In Fig. \ref{fig:235UStatistics}(a), we plot the average difference between the mean MDN prediction and the exact energy-dependent $Y(A)$ parameterization,
\begin{equation}
\overline{\Delta Y(A)} = \frac{1}{N} \sum \limits _{i=1}^{N} \left | \overline{Y^{MDN}}(A_i) - Y^{exact}(A_i) \right |,
\label{eqn:avgDeltaY}
\end{equation}

\noindent where $\overline{Y^{MDN}}(A_i)$ is the average MDN prediction, $Y^{exact}(A_i)$ is the energy-dependent parameterization, and $N$ is the number of masses, $A_i$, for which $Y(A)$ was calculated.  Within the energy range covered by the training set, with the exception of the lowest incident energies, $\overline{\Delta Y(A)}$ remains very close to zero at an almost constant value across the whole range of incident energies.  As the incident energy increases outside of the values included in the training set, the MDN mean values begin to deviate from the ground truth.  This deviation is seen both in panels (e) and (f) of Fig. \ref{fig:235U} and in Fig. \ref{fig:235UStatistics}(a).  It is also important to note that by constructing the training set without any information about multi-chance fission, the network has no knowledge of how the emission of pre-fission neutrons should should impact the mass yields as the incident energy increases.  A comparison of the predicted yields from the MDN and experimental yields above the second-chance fission threshold should show larger deviations than the comparison between the MDN yields and the yields labeled exact in Fig. \ref{fig:235U}, which only take into account first-chance fission where no neutron is emitted from the compound nucleus before fission. 


In addition, we can calculate the average uncertainty, $\varepsilon$, from Eq. (\ref{eqn:averageUncertainty}) for each of the MDN predictions in the testing set.  These values are shown in Fig. \ref{fig:235UStatistics}(b) as a function of the incident neutron energy.  Because the training set only included $Y(A)>10^{-3}$, $\varepsilon$ is also calculated with the same cut-off.  When the lower yields in the symmetric mass region are included, $\varepsilon$ increases by orders of magnitude; the MDN is not able to accurately predict these low yields, having never seen them in the training set.  Again, it appears that the scaling of the fission yields is extremely important to the training process.  When the yields below $10^{-3}$ are removed from the analysis, the average uncertainty on the mass yields only varies by about 6\% over the entire testing energy range.  It is also reassuring to see that the largest uncertainties are on the predicted yields that are farthest in energy from those energies included in the training set.

\section{Outlook}
\label{sec:outlook}

In this work, we explored the use of a probabilistic machine learning technique, the Mixture Density Network (MDN), to propagate uncertainties from training sets to predictions for fission fragment mass yields.  We showed that the MDN can reliably reproduce the level of accuracy on the training data set for mass yields of $^{252}$Cf(sf); although for training data with very small uncertainties (on the order of 1\%), the propagated uncertainties are inflated by about 10\%.  Without including any information on the physical constraints of the mass yields (e.g. normalization and symmetry), the MDN prediction spreads the distribution of the normalization values but keeps the symmetry between the heavy and light fragments within the spread of the training set.  When these constraints are explicitly built in to the training set through feature engineering, the normalization of the MDN mass yields remains the same but the symmetry between the heavy and light fission fragment yields improves.  In addition, we were able to extrapolate between and interpolate beyond training samples using energy-dependent mass yields for $^{235}$U(n,f) only training on yields for integer values of the incident neutron energy up to $E_\mathrm{inc}=4$ MeV.

The MDN appears to be a promising tool to propagate and quantify uncertainties on nuclear data without an explicit model.  There are still several topics that need to be investigated when using this algorithm.  The first is in the scaling of the training set, particularly when it spans several orders of magnitude.  Second, we would like to sample the training set from a covariance matrix between the mass in the yield distribution in order to include more of the correlations between masses directly.  In addition, when dealing with nuclear data, we are often in the situation where we will have partial data (e.g. knowing the most probable heavy and light masses but not the full yield distribution).  It will be crucial to be able to include this type of data in combination with full data sets or to treat this type of data when nothing else has been measured.  Towards this second goal, transfer learning is often used to more robustly fill in sparse data sets, when a second similar set is more robust.  This will become much more important as we move towards training on and predicting multi-dimensional observables, such as joint yields in mass, charge, and kinetic energy for several actinides simultaneously.  Finally, as with any newly applied technique, it would be useful to compare the uncertainties coming from the MDN to other more common uncertainty quantification techniques, such as standard $\chi^2$ minimization and covariance propagation or Monte Carlo Bayesian methods where there is an explicit functional form to fit to data.  

\ack
This work was performed under the auspice of the U.S. Department of Energy by Los Alamos National Laboratory under Contract 89233218CNA000001 and was supported by the Office of Defense Nuclear Nonproliferation Research \& Development (DNN R\&D), National Nuclear Security Administration, U.S. Department of Energy.  We gratefully acknowledge the support of the U.S. Department of Energy through the LANL/LDRD Program and the Center for Non Linear Studies.

\newpage
\normalsize
\bibliography{MDN}

\end{document}